\pdfoutput=1
\documentclass[a4paper,fleqn,usenatbib]{mnras}
\usepackage[T1]{fontenc}
\usepackage{ae,aecompl}
\usepackage{psfig}  
\usepackage{graphicx} 
\usepackage{dcolumn}  
\usepackage{bm}       
\usepackage{amssymb,amsmath}  
\usepackage{color,colortbl}
\usepackage{float}
\usepackage{nicefrac}

\newcommand{\etal}{{\it et al.\,}}
\newcommand{\cola}{{\tt{COLA} }}
\def\Msun{\, h^{-1} \, {\rm M_{\odot}}}
\newcommand{\hompc}{\,h\,{\rm Mpc}^{-1}}
\newcommand{\mpcoh}{\,h^{-1}\,{\rm Mpc}}
\definecolor{Gray}{gray}{0.94}

\begin{document}

\title[ICE-COLA]
{ICE-COLA: Towards fast and accurate synthetic galaxy catalogues optimizing a quasi $N$-body method}
\author[Izard \etal]{
  Albert Izard, Martin Crocce, \& Pablo Fosalba \\
Institut de Ci\`encies de l'Espai, IEEC-CSIC, Campus UAB, Carrer de Can Magrans, s/n,  08193 Bellaterra, Barcelona, Spain}

\twocolumn   
\maketitle 

\begin{abstract}
Next generation galaxy surveys demand the development of massive ensembles of galaxy mocks to model the observables and their covariances, what is computationally prohibitive using $N$-body simulations. \cola is a novel method designed to make this feasible by following an approximate dynamics but with up to 3 orders of magnitude speed-ups when compared to an exact $N$-body. In this paper we investigate the optimization of the code parameters in the compromise between computational cost and recovered accuracy in observables such as two-point clustering and halo abundance. We benchmark those observables with a state-of-the-art $N$-body run, the MICE Grand Challenge simulation (MICE-GC). We find that using 40 time steps linearly spaced since $z_i \sim 20$, and a force mesh resolution three times finer than that of the number of particles, yields a matter power spectrum within $1\%$ for $k \lesssim 1 \hompc$ and a halo mass function within $5\%$ of those in the $N$-body. In turn the halo bias is accurate within $2\%$ for $k \lesssim 0.7 \hompc$ 
whereas, in redshift space, the halo monopole and quadrupole are within $4\%$ for $k \lesssim 0.4 \hompc$. These results hold for a broad range in redshift ($0 < z < 1$) and for all halo mass bins investigated ($M > 10^{12.5} \Msun$). To bring accuracy in clustering to one percent level we study various methods that re-calibrate halo masses and/or velocities. We thus propose an optimized choice of \cola code parameters as a powerful tool to optimally exploit future galaxy surveys.
\end{abstract}


\section{Introduction}

Present and planned galaxy surveys like DES\footnote{\texttt{http://www.darkenergysurvey.org/}}, LSST\footnote{\texttt{http://www.lsst.org}}, Euclid\footnote{\texttt{http://www.euclid-ec.org/}}, eBOSS\footnote{\texttt{http://www.sdss.org/surveys/eboss/}}, DESI\footnote{\texttt{http://desi.lbl.gov/}}, will generate a wealth of high-quality data that will allow to test the nature of dark-energy and constrain possible deviations from the standard cosmological model based on General Relativity \citep{Weinberg13}.

An optimal extraction of cosmological parameters from those very large and complex datasets will ultimately rely on our ability to model cosmological observables 
and their covariances with high accuracy. This entails the development of synthetic observations based on mock catalogues produced from numerical simulations. The requirement of sampling large cosmological volumes while still resolving small scales is a big challenge to current $N$-body simulation codes (\citealt{Kim11,Angulo12,Alimi12,Skillman14,Heitmann14,MICEI}; for a review see \citealt{Kuhlen12}). Moreover, hundreds or thousands of realizations are needed for robustly estimating covariance matrices (crucial for cosmological parameter estimation, see \citealt{Taylor13,Blot15}) or for propagating errors in complex and non-linear analysis (e.g. BAO reconstruction, see \citealt{Takahashi09,Kazin14,Ross15}). Yet, producing massive ensembles of $N$-body mocks is computationally prohibitive and alternative routes need to be devised in order to face the enormous challenge.

A convenient approach to reproduce the observed galaxy distribution on large scales is to populate dark matter haloes with galaxies using either Semi-Analytic Models or techniques relying on the Halo model, such as the Halo Occupation distribution or the Sub-halo Abundance Matching \citep[see][for a comparison of models]{Knebe15}. Some of these techniques do not need to fully capture internal halo substructure since they only use reliable positions and mass estimates for haloes, which can be predicted by approximate methods. A simplified and cheaper evolution of the density field is enough to approximately predict the abundance and clustering of collapsed regions to build halo mock catalogues. Available methods that are based on this idea include: \textsc{pinocchio} (PINpointing Orbit-Crossing Colapsed HIerarchical Objects,
\citealt{Monaco02}; \citealt{Monaco13}), PThaloes (\citealt*{Scoccimarro02}; \citealt{Manera15}) and recently \cola (COmoving Lagrangian Acceleration, \citealt*{Tassev13,Tassev15}).

An alternative approach is to use prescriptions to assign haloes in a density field produced by a simple gravity solver. This is the case of the log-normal model \citep{Coles91}, QPM \citep*[Quick Particle Mesh][]{White13}, \textsc{patchy} \citep[PerturbAtion Theory Catalog generator of Halo and galaxY distributions][]{Kitaura15}, EZmocks \citep[effective Zel'dovich approximation mock catalogues][]{Chuang15b}. Thus these methods constitute a more direct modeling for the galaxy distribution. The drawback is that they contain many internal parameters describing properties such as the galaxy clustering and abundance that have to be fit in order to correctly reproduce observations. \cola can be categorized as a semi-$N$-body method and therefore has higher computational requirements than other methodologies. But is more predictive and yields accurate high order clustering statistics. For a recent performance comparison of fast mock methods see \citet{Chuang15a}.

A common feature in most of the fast methods is that they evolve mass particles using Lagrangian Perturbation Theory (LPT). \cola is unique because on top of the analytical trajectory it adds a residual displacement computed by an $N$-body solver. Equations of motion are solved in a frame comoving with LPT observers which, at a given perturbative order, encodes more non-linear growth information than the corresponding Eulerian approach. This guarantees an accurate description of the dynamics on large scales where the evolution is quasi-linear. The numerical evolution is simplified with respect to full $N$-body codes, making use of a fine Particle-Mesh (PM) algorithm, and evaluating forces for a few (i.e, order of ten) time steps. haloes can then be identified running a finder in the evolved dark matter particle distribution, in the same way is done for an $N$-body.

In {\tt COLA}, the displacement field $\bmath{x}$ is decomposed into two terms: $\bmath{x}_{\rmn{LPT}}$ describes the LPT trajectory and $\bmath{x}_{\rmn{res}}$ is the residual displacement with respect to the LPT path,
\begin{eqnarray}
\bmath{x}_{\rmn{res}}(t) \equiv \bmath{x}(t)-\bmath{x}_{\rmn{LPT}}(t)
\label{eq:cola_displ_field}
\end{eqnarray}

The equation of motion in a pure gravitational simulation relates the acceleration to the Newtonian potential $\Phi$: $\partial_{t}^{2} \bmath{x}(t) = -\nabla\Phi(t)$. Using Eq. \ref{eq:cola_displ_field}, it can be rewritten as
\begin{eqnarray}
\partial_{t}^{2}\bmath{x}_{\rmn{res}}(t) = -\nabla\Phi(t)-\partial_{t}^{2}\bmath{x}_{\rmn{LPT}}(t),
\end{eqnarray}
where $\Phi$ is evaluated at the position $\bmath{x}$. \cola discretizes the time derivatives only on the left-hand side, while uses the LPT expression at the right-hand side. More recently, \citet{Tassev15} extended this reformulation to the Eulerian domain, which allow the simulation of a sub-volume embedded in a larger effective simulation box.

\citet{Kazin14} developed a parallel version of {\tt COLA}, suitable for the massive production of mock catalogues, which they used for constraining the BAO signal in the WiggleZ survey. This paper is based on their implementation.

\cola is able to accurately follow the evolution of the dark matter distribution on scales of few Mpcs. More challenging is to reproduce the birth and growth of haloes, which display high density contrasts and non-linear dynamics sustained by virialisation. Halo formation is very sensitive to the degree of approximation in the dynamics at small scales and a minimum accuracy is indispensable to generate reliable halo mock catalogues. Therefore, it is essential to assess the performance of the method under different values for the internal code parameters that describe the spatial and temporal discretization of the gravitational evolution.

In this paper we develop a suite of large \cola simulations to explore the impact of varying internal code parameters on basic observables such as the matter real-space power spectrum and the halo mass function. In particular, we explore the size of the force evaluation mesh, the number and distribution of time steps, and the initial redshift, in combination with mass resolution. As a reference we use the $N$-body simulation MICE-GC, which has been extensively validated \citep{MICEI,MICEII,MICEIII}. Within this parameter space we find the configuration that yields the best accuracy in power spectrum and mass function (for a range of halo masses, comoving scales and redshifts) without a large increase in computational cost. For the optimal parameters, we also characterize the recovered accuracy for halo clustering in real and redshift space. 
The above procedure yields what can be regarded as the optimal accuracy of the code on its own. To improve it further one needs to rely on 
simple corrections to halo masses using an external simulation.
In \citet{Tassev13} they already showed that the halo mass should be corrected and they used a constant multiplicative factor at all masses and redshifts. In this work, the first correction we explore aims at matching the halo abundance
and the second the halo clustering amplitude on large scales. Thanks to them, deviations in halo clustering can be reduced to within the percent level in most situations. 

This paper is organized as follows. In Sec. \ref{sec:simulations} we describe the simulations used throughout this paper, as well as their analysis. Then we start in Sec. \ref{sec:10_easy_steps} discussing the capabilities and limitations of \cola when run with as few as 10 time steps. Sec. \ref{sec:optimization} presents the code parameters exploration, where we show the accuracy that can be achieved using different configurations and we give optimal parameters. The next two sections use runs with optimal code parameters. In particular, Sec. \ref{sec:halo_clustering} shows the results for halo clustering and how they can be improved by re-calibrating halo masses. In Sec. \ref{sec:z-space} we test the performance of \cola in redshift space. We summarize and discuss our findings in Sec. \ref{sec:conclusions}. 

\section{Simulations}
\label{sec:simulations}

\subsection{MICE-GC: the benchmark $N$-body run}
\label{sec:micegc}

In this paper we use a full $N$-body run to benchmark the results of {\tt COLA}: the MICE Grand Challenge (MICE-GC)\footnote{More information is available at \texttt{http://www.ice.cat/mice}.} simulation. This simulation and its products have been extensively validated, the dark-matter and halo outputs are described in \citet{MICEI} and \citet{MICEII}. In addition, lensing maps are described in \citet{MICEIII} while \citet{Carretero15} and \citet{Hoffmann15a} detail the HOD implementation used to produce galaxies mocks and the higher-order clustering, respectively. MICE-GC evolved $4096^3$ particles in a volume of $(3072\mpcoh)^3$ using the \textsc{gadget-2} code \citep{Springel05} assuming a flat $\Lambda$CDM cosmology with $\Omega_m=0.25$, $\Omega_{\Lambda}=0.75$, $\Omega_b=0.044$, $n_s=0.95$, $\sigma_8=0.8$ and $h=0.7$. This results in a particle mass of $2.93\times10^{10}\Msun$. The initial conditions were generated at $z_i=100$ using the Zel'dovich approximation 
and a linear power spectrum generated with  \textsc{camb}\footnote{\texttt{http://camb.info}}. We make use of dark matter and halo catalogues of comoving outputs at $z=0, 0.5$, $1.0$ and $1.5$.

Haloes were identified using a Friends-of-Friends (FoF) algorithm \citep{Davis85} with a linking length of 0.2.

It is known that long-lived transients from the initial conditions affect the abundance of massive haloes and the clustering towards small scales, even for high starting redshifts if the Zel'dovich approximation is used \citep{1998MNRAS.299.1097S,Crocce06}. We have performed a set of dedicated \textsc{gadget-2} $N$-body simulations to investigate and correct such effects.
This is discussed in Appendix \ref{sec:transients_correction}. Corrections are $\lesssim 2\%$ for 2-pt matter clustering and $2-5\%$ for halo abundance on the regime and redshifts of interest. 
Hence whenever we compare to matter power spectra and mass functions in MICE-GC, we take into account such corrections.
We find that the transient effects are below the 1 per cent level for halo clustering, so we consider that we can safely neglect the correction for those measurements.

\subsection{Parallel  {\tt COLA} code}
\label{sec:cola_code}

We use a parallel version of \cola developed by J. Koda for the WiggleZ survey \citep{Kazin14}, and kindly made available to us. It includes both the generation of random Gaussian initial conditions using 2LPT \citep{Crocce06} and the FoF halo finder \citep{Davis85} running on the fly. Particle forces are computed using a Particle Mesh (PM) code with a grid that is finer than the mean inter-particle distance. We shall refer to this as the ${\rm PM}_{grid}$ factor, which \citet{Tassev13} suggest to set to 3 in order to adequately resolve small mass haloes, meaning that in total there are $3^3$ more grid cells than particles (the total number of cells in the grid is $\mathrm{number \,particles} \times {\rm PM}_{grid}^3$). At each time step, four fast Fourier transforms (FFT)\footnote{The code relies on the FFTW package in its MPI version for distributed memory parallelization (\texttt{http://www.fftw.org/}).} are needed: one to convert the density field to Fourier space and three for transforming back to real space the forces in each of the spatial dimensions. We have added some new capabilities to the code such as the generation of matter power spectra on-the-fly and the parallel storage of the matter density field interpolated onto large mesh grids (in particular we use $1024^3$ cells using the Cloud-in-Cell assignment).

\subsection{Simulation suite}
\label{sec:simulation_suite}

In order to investigate optimal code parameter set ups we implemented several \cola runs. 
They were performed on the MareNostrum supercomputer at BSC\footnote{Barcelona Supercomputing Center (\texttt{http://www.bsc.es}).}. We used the same cosmological model, the same linear power spectrum and same particle mass as MICE-GC (except in cases where we are interested in exploring mass resolution effects). Most of the runs used $2048^3$ particles in a box size of $1536\mpcoh$, that is, a factor 8 smaller volume than MICE-GC. It provides nonetheless a very large cosmological volume, but for some particular runs of interest we have produced additional realizations in order to reduce sampling variance. A typical run uses 1024 cores, with maximum memory consumptions of 2.6Tb for ${\rm PM}_{grid}=3$ and takes around 40 minutes wall-clock time for 40 time steps. This means that the CPU time consumed is less than 1 khour, to be compared with the 3 Mhours that needed MICE-GC having 8 times more particles, which gives a speed-up factor between two and three orders of magnitude with respect to a full $N$-body simulation with the same number of particles.

The code parameters we have used for the runs are listed in Table \ref{table:cola_runs}. We have varied the ${\rm PM}_{grid}$ factor, the number of time steps, the time sampling and the initial redshift. Additionally, we explore the effect of decreasing the mass resolution by decreasing the number of particles while keeping the box size constant. We also reduce the box size while keeping the particle load constant for a better mass resolution. All runs use the same seed number for the generation of the initial conditions (except for those which add more realizations to the same parameter configuration), what cancels out cosmic variance between different simulations (but not with respect to MICE-GC, which uses a different box size). In what follows, we shall assume a default set of parameters, unless stated otherwise: $2048^3$ particles, $L_{box}=1536\mpcoh$, ${\rm PM}_{grid}=3$ and a distribution of time steps linear in the scale factor.

\begin{table*}\normalsize 
\begin{tabular}{ccccccc}

\hline \\

$N_{{\rm realizations}}$ & \parbox[c]{1cm}{Particle\\number}  & $L_{box}$  &  ${\rm PM}_{grid}$  & $N_{{\rm steps}}$  &  $z_{{\rm i}}$  & \parbox[c]{1.6cm}{Time steps\\distribution}  \\
\\
\hline \\

$1$  &  $\underline{1024}^3$  &  $1536$  &  $3$  &  $10$  &  $9$  &  $\propto a$   \\

$1$  &  $\underline{1536}^3$  &  $1536$  &  $3$  &  $10$  &  $9$  &  $\propto a$   \\

$1$  &  $2048^3$  &  $1536$  &  $\underline{2}$  &  $10$  &  $9$  &  $\propto a$   \\
$2$  &  $2048^3$  &  $1536$  &  $3$  &  $10$  &  $9$  &  $\propto a$   \\

$1$  &  $2048^3$  &  $1536$  &  $3$  &  $\underline{20}$  &  $9$  &  $\propto a$   \\

$1$  &  $2048^3$  &  $1536$  &  $3$  &  $\underline{40}$  &  $9$  &  $\propto a$   \\

$1$  &  $2048^3$  &  $1536$  &  $\underline{2}$  &  $40$  & $19$  &  $\propto a$   \\
\rowcolor{Gray}
$48$ &  $2048^3$  &  $1536$  &  $3$  &  $40$  & $19$  &  $\propto a$   \\

$1$  &  $2048^3$  &  $\underline{768}$  &  $3$  &  $40$  & $19$  &  $\propto a$    \\

$8$  &  $2048^3$  &  $1536$  &  $3$  &  $40$  & $\underline{39}$  &  $\propto a$   \\

$1$  &  $2048^3$  &  $1536$  &  $3$  &  $40$  & $\underline{39}$  & $\underline{\propto \log a}$ \\

$1$  &  $2048^3$  &  $1536$  &  $3$  &  $40$  & $\underline{39}$  & $\underline{\propto a^{0.8}}$ \\

$1$  &  $2048^3$  &  $1536$  &  $3$  &  $40$  & $\underline{100}$ &  $\propto a$   \\

	 \\
\hline
\end{tabular}
\caption{Code parameters used in this paper. The box sizes are in units of $\mpcoh$. We underline parameter values that are distinctive for a run and the highlighted row corresponds to the optimal setup, which provides the best accuracy. For the latter, each realization needed 1024 cores during 40 minutes and 2.6Tb of memory. In Sec. \ref{sec:scaling_computational_requirements} we discuss how to extrapolate those computational requirements to other parameter configurations. The total number of cells in the force mesh is: $\mathrm{number\,particles \times {\rm PM}_{grid}^3}$}
\label{table:cola_runs}
\end{table*} 

\subsection{Scaling of computational requirements}
\label{sec:scaling_computational_requirements}

We explain in this section how the run-time and memory consumption of a single \cola realization scale with code parameters, so that combined with the information provided in the previous subsection one can extrapolate the numbers to other configurations.

The initial redshift and the time sampling distribution have no effects at all on the computational requirements. The run-time is largely dominated by the computations of FFTs during force evaluation at each step. For our default configuration and 40 time steps, the code spends only 10 per cent of the time in computations not related to the PM algorithm, such as the initial set up and I/O. For this reason, the run-time increases roughly linearly with the size of the FFT and the number of time steps. Such transforms scale as $\propto \mathcal{O}(n\log n)$, where $n$ is the number of grid points, and since these are proportional to ${\rm PM}_{grid}^3$ we have that the run-time scales roughly as $\propto {\rm PM}_{grid}^3$ (for large numbers we can neglect the $\log n$ factor).

Given a constant number of particles, the memory consumption depends only on the size of the force mesh. The allocation of memory from the PM part represents around 60 per cent for ${\rm PM}_{grid}=3$ and it scales as ${\rm PM}_{grid}^3$.

\subsection{Power spectrum and mass function measurements}
\label{sec:analysis}

We determine the matter and halo power spectra interpolating the particles into a cubic grid of $1024^3$ cells via a Cloud-in-Cell (CiC) assignment.
We then obtain the density in k-space by doing a FFT and estimate the band power by averaging the square over the range of modes corresponding to a k-bin: $P(k)=\langle \left| \delta_{\bmath{k}} \right|^2 \rangle$. The mass assignment into a finite size grid introduces a filtering artifact that we compensate by deconvolving the CiC window function, which in Fourier space is simply a division. We correct for aliasing effects due to the finite sampling as in \citet{Jing05}. Lastly, monopole measurements of halo power spectrum are corrected for shot-noise assuming a Poisson noise.

For haloes, we restrict the analysis to those having more than 100 particles, corresponding to $M\ge10^{12.5}\Msun$ for most of the runs. Halo masses are defined using the Warren correction \citep{Warren06}: $M=m_p N(1-N^{-0.6})$, where $m_p$ is the particle mass and $N$ the number of particles. This is irrelevant for most of the runs that use the same mass resolution as MICE-GC, but provides better agreement for lower mass resolution runs. We build three halo samples according to the mass cuts listed in Table \ref{table:mass_bins}. Mass function measurements contain error bars estimated by Jack-knife re-sampling using 64 different cubic sub-volumes and only mass-bins whose relative error is less than 5\% are shown (see e.g. Fig. \ref{fig:10steps}).

\begin{table}\normalsize  
\centering
\begin{tabular}{cc}

Sample  &  Mass range $[\log (M/\Msun)]$ \\

\hline

M1  &  $12.5 - 13.0$ \\
M2  &  $13.0 - 13.5$ \\
M3  &  $\ge13.5$ \\

\end{tabular}
\caption{Halo mass samples used throughout this paper, defined by halo mass, at $z=0,0.5$ and $1$.}
\label{table:mass_bins}
\end{table} 

\section{Limitations of 10 time steps}
\label{sec:10_easy_steps}

The \cola method is designed to use very few time-steps, so that a high speed-up of more than two orders of magnitude with respect to a full $N$-body is achieved \citep{Tassev15}. 
In this section we explicitly test the accuracy, as a function of scale and redshifts, of the original \cola configuration of 10 time steps with scale factor linearly distributed between redshift 9 and 0 and a ${\rm PM}_{grid}$ factor of 3.

\begin{figure} 
\includegraphics[width=0.99\columnwidth,trim={0 6.5cm 0 0},clip]{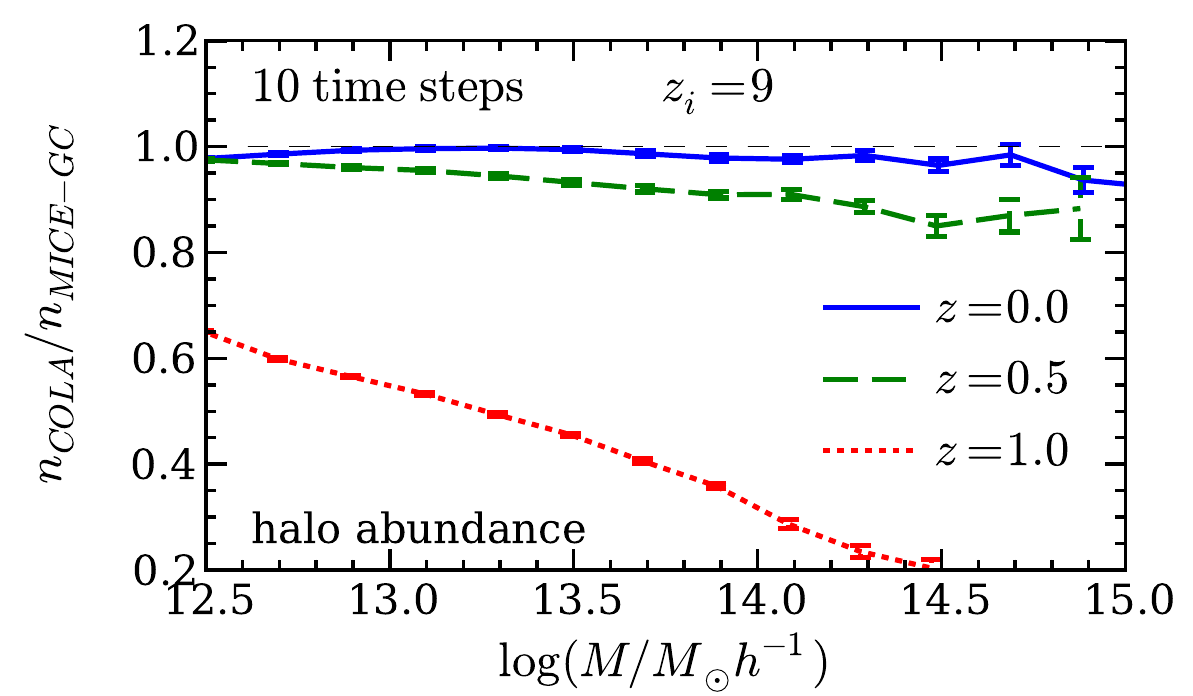}
\caption{Mass function when using \cola with 10 time steps starting at $z_i=9$. Solid, dashed and dotted lines correspond to $z=0$, 0.5 and 1 respectively. At high redshift there is a large discrepancy that is partially solved at $z=0.5$.}
\label{fig:10steps}
\end{figure}

With only 10 time steps, the matter power spectrum is accurate at the 5\% level to $k\sim0.5\hompc$ (see Sec. \ref{sec:nsteps}). This allows the exploration of non-linear scales, but only to some limited extent. At the halo level, however, the situation is more complex. Figure \ref{fig:10steps} shows that the mass function is severely underestimated at $z=1$, especially at large masses. The problem is not so visible at $z=0.5$, where the disagreement is at most at the $15\%$ level and at $z=0$ it is within the $5\%$ for all masses. Likewise, we have checked that the halo bias is overestimated by as much as 20\% at $z=1$ and agrees within few percent level at lower redshifts. Both effects (underestimation of the mass function and overestimation of the halo bias) can be explained by a halo mass underestimation at high redshift, when the evolution has been computed with very few time steps.

We suggest that these problems could come from a higher difficulty of achieving relaxation inside haloes when the time discretization is too coarse. Particles evolve according to the mean gravitational potential that arises from the smooth distribution, but are also affected by individual encounters. The relaxation time is related to the moment when the latter start to significantly contribute to the dynamics, boosting the re-distribution of kinetic energy and achieving a dynamical equilibrium in the system \citep{Dehnen11}.

Each time step is a chance for particles to interact with each other, but if we reduce them drastically the re-distribution of energy is unphysical suppressed. This is critical for those haloes that have not relaxed yet. Since the relaxation time is proportional to the number of particles of a halo \citep{Binney-Tremaine}, the effect is larger for high mass haloes. The halo formation time increases and merging processes are poorly captured, producing halo masses largely underestimated for  $z>0$, before 10 time steps have been completed. Note that full $N$-body codes with adaptive time steps schemes trigger finer time samplings at high density regions and halo formation is properly tracked.

\begin{figure*} 
\includegraphics[width=0.69\columnwidth]{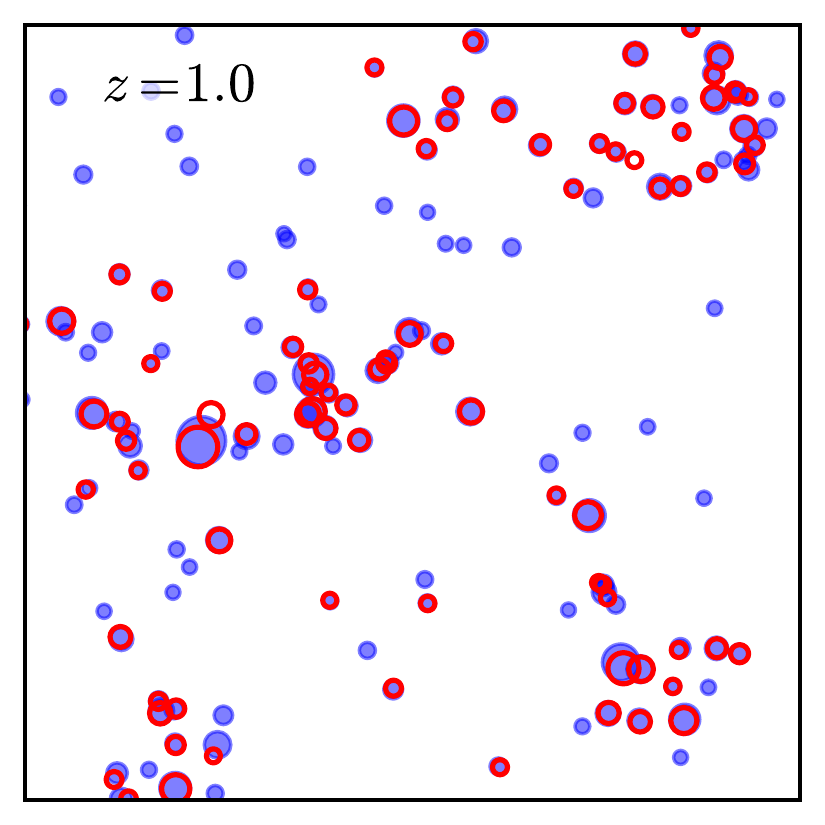}
\includegraphics[width=0.69\columnwidth]{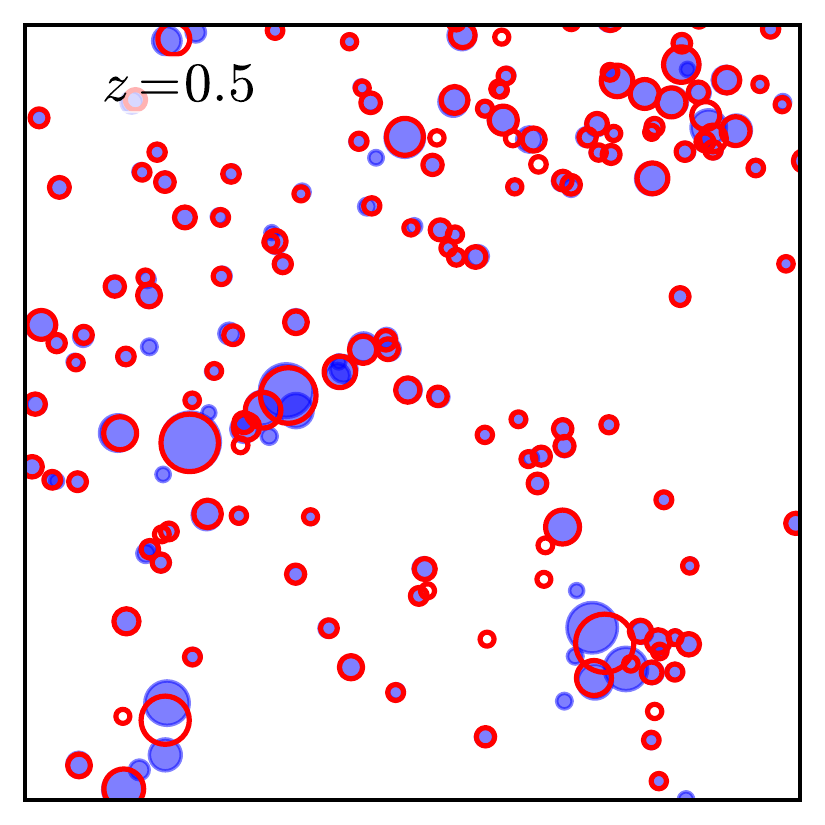}
\includegraphics[width=0.69\columnwidth]{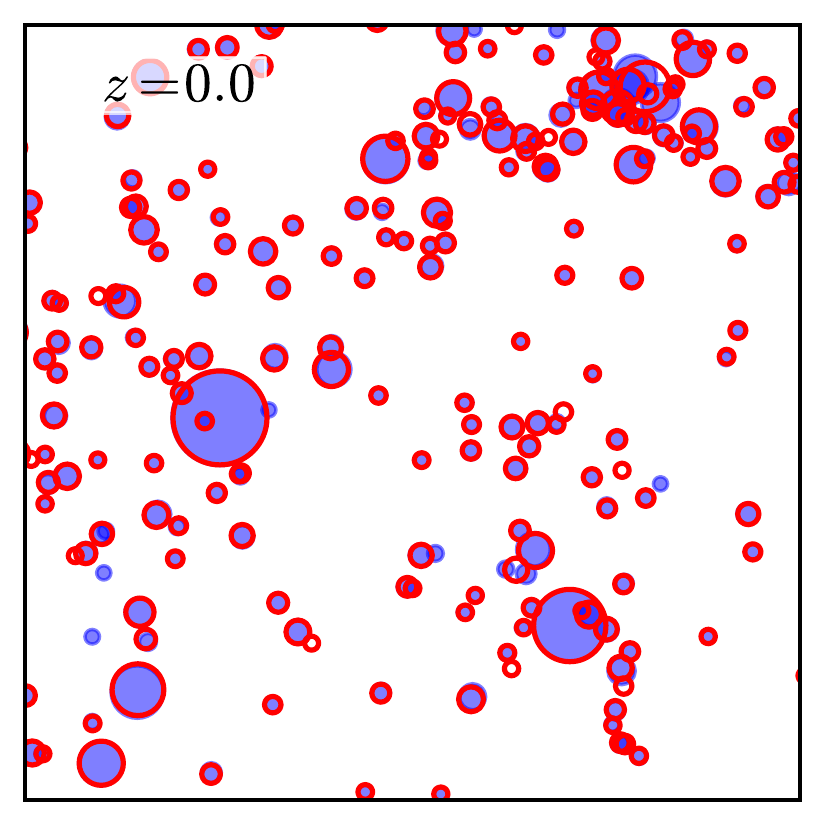}
\caption{Spatial distribution of haloes in two different \cola runs that differ only in the number of time steps: 10 are represented by open circles and 40 by filled ones. Different panels show redshifts 1.0, 0.5 and 0 from left to right. The slices have a width of 50$\mpcoh$ and are 25$\mpcoh$ thick. The radii of circles are proportional to $M^{1/3}$ and match the $r_{200}$ values, so that they reflect the physical size of haloes. Only those with more than 200 particles are shown (which corresponds to $6\times10^{12}\Msun$). The largest halo at $z=0$ has a mass of $1.87\times10^{15}\Msun$ for a run with 40 time steps and a $4.5\%$ less for 10 time steps. At $z=1$ the matching between the runs is poor, with the abundance under-estimated by $50\%$ or more with 10 time-steps. The agreement improves as one approaches $z=0$.}
\label{fig:halo_distr}
\end{figure*} 

To visually confirm this idea, we show in Fig. \ref{fig:halo_distr} the halo distribution of two runs with 10 and 40 time steps in red open and blue filled circles, respectively. The initial conditions and the rest of parameters are kept the same, so that differences are due only to the number of time steps. In the left panel, we confirm that at $z=1.0$ massive haloes are in general not properly tracked with 10 time steps and they seem to appear fragmented as smaller mass haloes. And not all of the low mass haloes are identified. Nevertheless, in the right panel at $z=0$, when 10 time steps have been completed, the agreement is much better on halo masses, positions and abundance at all masses. This visual inspection suggests that one needs $\ga10$ time steps before halo properties converge at the redshift of interest.

We find that relaxation effects when only 10 time steps are used can be reduced using a higher particle mass (e.g. above $10^{11}\Msun$). The number of particles of the haloes decreases and, therefore, their relaxation time as well. We checked that disagreements in the mass function and halo clustering are indeed lower, but this apparent improvement is however lost for other statistics where mass resolution is important.

Evolving particles with just ten time steps before the redshift of interest, therefore, provides accurate results only at large scales ($k\le0.3\hompc$) and low redshifts. We might have stronger requirements that clearly push to go beyond 10 time steps to surpass those limitations.

\section{Optimization}
\label{sec:optimization}

A gravity solver algorithm discretizes both temporal and spatial dimensions (and the mass as well) in order to numerically solve for the dynamical evolution. The idea behind \cola is to reduce numerical computations as much as possible while still capturing the growth of structure on large scales. This can be controlled with few internal code parameters: the number of time steps, the time sampling distribution, the initial redshift and the size of the force mesh grid, in combination with the mass resolution and/or the box size. Note however that \cola is not fundamentally different from a full $N$-body in the sense that as one increases the requirements on such parameters the numerical integration of particle trajectories becomes more accurate and \cola would eventually converge to a full $N$-body.

In this section we explore the code parameter space in order to understand their impact on observables and determine regions that provide optimal results in terms of accuracy versus running time (or memory usage). We assess the performance by comparing the \cola dark matter power spectra up to $k \sim 1\hompc$, and halo mass functions for $M \gtrsim 10^{12.5}\Msun$, against those in the reference $N$-body run. A key difference from previous works \citep{Kazin14,Howlett15,Leclercq15} is that we aim at reproducing those observables simultaneously across a large redshift range ($0 \le z \le 1$). As we will see next, the requirements in this case are more stringent that those needed for a single redshift output or halo mass bin.

\subsection{Number of time steps}
\label{sec:nsteps}

\begin{figure} 
\includegraphics[width=0.99\columnwidth]{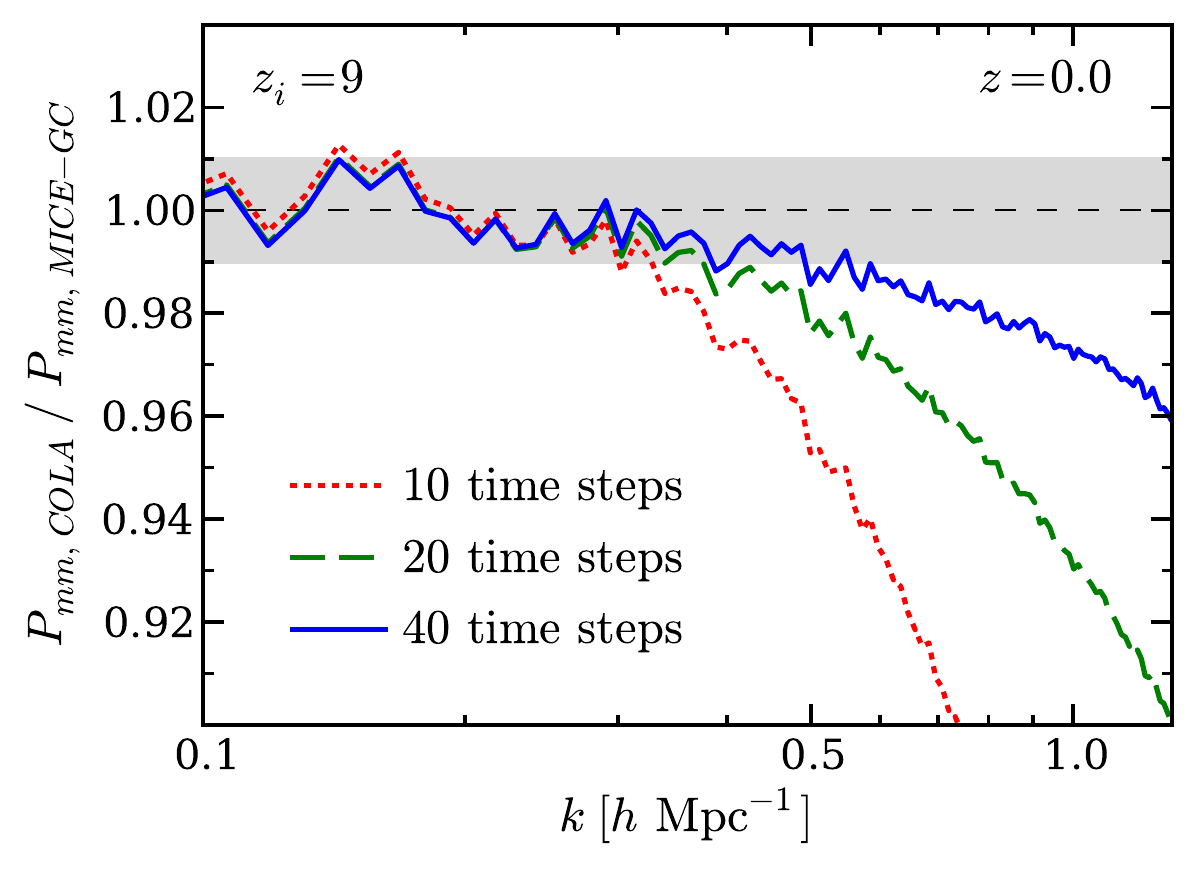}
\includegraphics[width=1.02 \columnwidth]{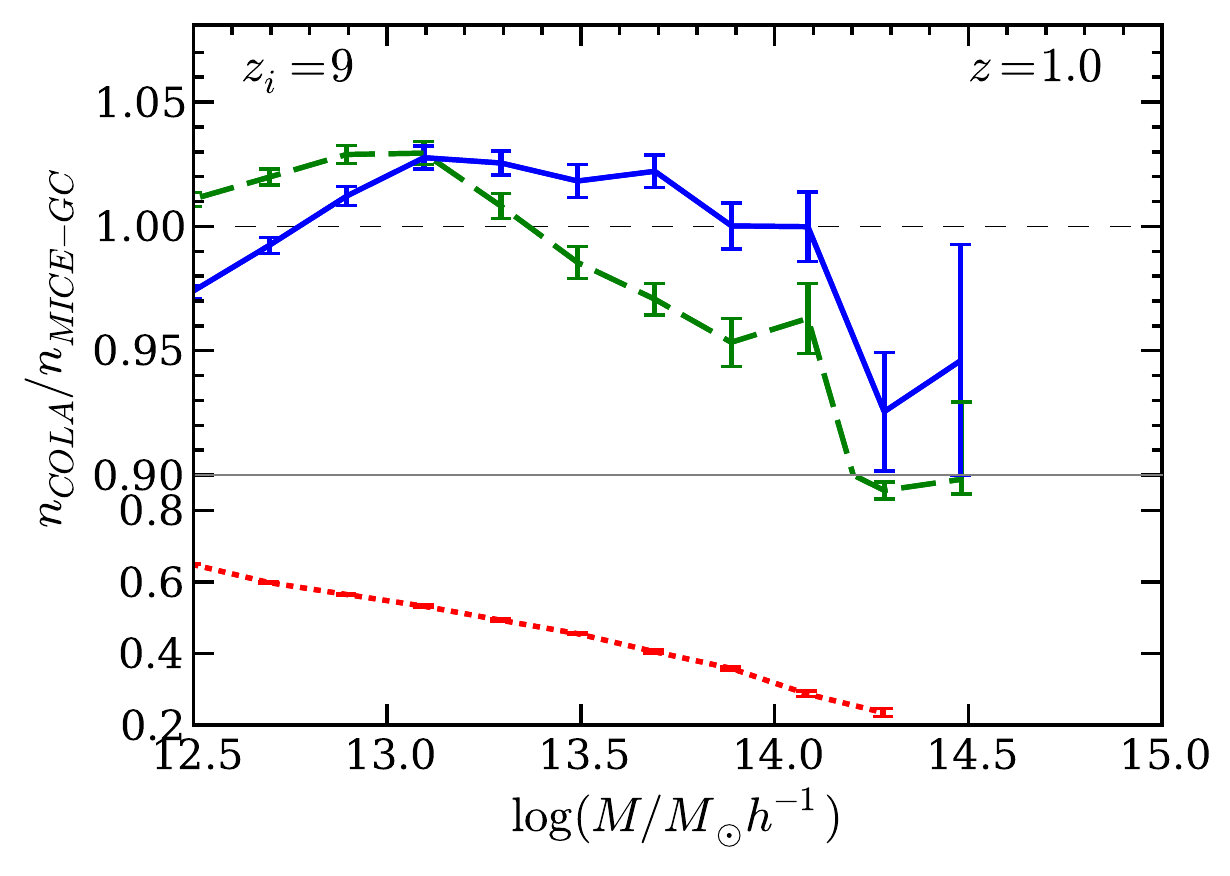}
\caption{Matter power spectrum in real space at $z=0$ (upper panel) and mass function at $z=1.0$ (lower panel) for three different choices for the number of time steps: 10, 20 and 40 in dotted, dashed and solid lines, respectively. The initial redshift is fixed at $z_i=9$. An increase in the number of time steps directly translates into a better accuracy at small scales. As for haloes, a high number of time steps is necessary to correctly predict the mass function at high redshift.}
\label{fig:pkmm_vs_nsteps}
\end{figure} 

The first parameter we vary is the number of time steps, with the initial redshift fixed to 9. The upper panel of Fig. \ref{fig:pkmm_vs_nsteps} displays the $z=0$ matter power spectrum for \cola runs with increasing number of time steps divided by the one measured in MICE-GC. The characteristic scale at which the ratio deviates from unity is progressively shifted towards higher wave-numbers as more time steps are included: there is a 2 per cent agreement up to scales of $k\sim$ 0.4, 0.6 and 0.8$\hompc$ for 10, 20 and 40 time steps respectively. This can still be improved adjusting the initial redshift according to the number of time steps (see Sec. \ref{sec:intial_redshift}). In particular, we check that doubling the number of time steps almost doubles the characteristic wavenumber where the power spectrum is significantly underestimated. This in turn means that there is room for higher accuracies with more than 40 time steps, although presumably the force mesh resolution would then soon become a limiting factor. We also find that these results for the matter power spectrum are to a good extent independent of the redshift analyzed.

However, for the mass function there is a higher sensitivity to the number of time steps at high redshift. As shown in the lower panel of Fig. \ref{fig:pkmm_vs_nsteps}, the large underestimation of the mass function at $z=1$ is solved by doubling the total number of time steps. With 20 time steps, 10 of them are computed before the redshift of interest. The abundance at high masses further increases by $\sim5$ per cent when moving from 20 to 40 time steps and the mass function is consistent with the one from MICE-GC.

At redshift 0 and 0.5 the mass function also increases with the number of time steps at masses above $\sim10^{13.5}\Msun$, although more moderately. Moving from 10 to 20 time steps, the mass function augments by $5-10\%$ at the high mass range and from 20 to 40 by $\la5\%$. At low masses, differences remain within 1 per cent for 20 and 40 time steps. We conclude, therefore, that the low mass regime of the mass function converges for 20 time steps but that 40 are necessary for the most massive haloes. In Appendix \ref{sec:pmonly} we show that, even with as many steps as 20 or 40, the 2LPT contribution in the \cola method is still key to achieve accurate results, as compared to PM only simulations.

\subsection{Time sampling distribution}
\label{sec:step_distribution}

The scale factor $a$ is the variable used to discretize the temporal axis in regular time steps. For that, we can choose a time sampling function $f(a)$ and distribute $n$ steps in intervals of constant width $\Delta f(a)$,

\begin{eqnarray}
\Delta f(a) =\frac{f(a_f)-f(a_i)}{n-1},
\end{eqnarray}

where $a_i$ is the initial scale factor and $a_f\equiv 1$ the final. If the resulting $\Delta a << 1$ then $\Delta a \approx [f'(a)]^{-1} \Delta f(a)$. For the linear case, we simply have $f(a)=a$ and the step width $\Delta a$ is constant. We can define the step density as the inverse of the step width: $\rho\equiv\nicefrac{1}{\Delta a}$. Since $\Delta f(a)$ is constant, then $\rho \propto f'(a)$.

\begin{figure} 
\includegraphics[width=0.99\columnwidth]{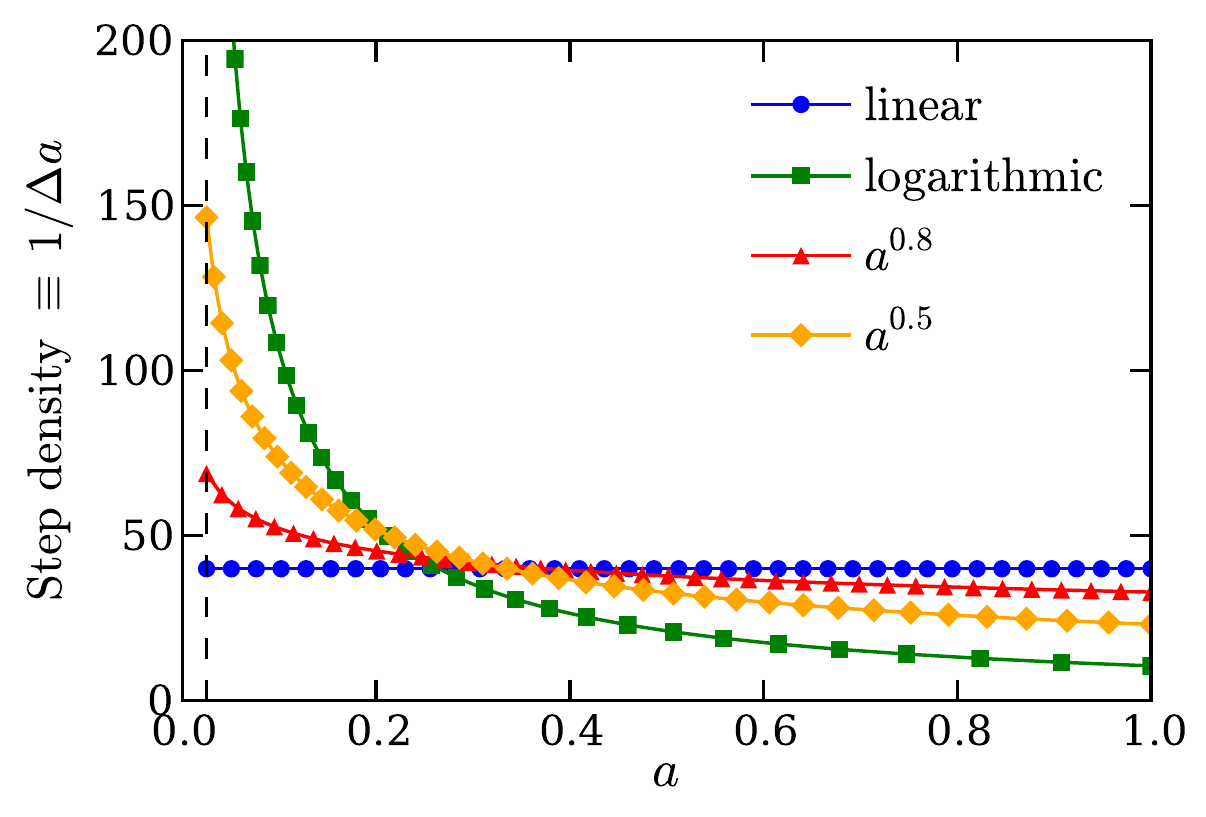}
\caption{Step density (or the inverse of the step width) as a function of the scale factor for different schemes of time sampling: linear (circles), logarithmic (squares) and using a power of the scale factor, where for the latter we show two cases with exponents 0.5 (diamonds) and 0.8 (triangles). In all cases we distribute 40 steps between $z_i=39$ and $z=0$ and the markers are located at the position of each step.}
\label{fig:time-stepping}
\end{figure}

In Sec.~\ref{sec:nsteps} we showed for the linear case how increasing the step density (which in that case is only set by the number of time-steps) improves the accuracy of the simulation, but at the expense of a higher computational cost.  In this section we explore which function $f$, and the corresponding step density $\rho \propto f'(a)$, produce a step distribution that is balanced in terms of accumulated errors over time. In Fig. \ref{fig:time-stepping} we show the step density for four different choices of the time sampling function. We distribute 40 time steps between $z_i=39$ and $z=0$ using a linear (circles), logarithmic (squares) or power law function $f(a)=a^p$ (diamonds and triangles for $p=0.5$ and $p=0.8$ respectively).

\begin{figure} 
\includegraphics[width=0.99\columnwidth]{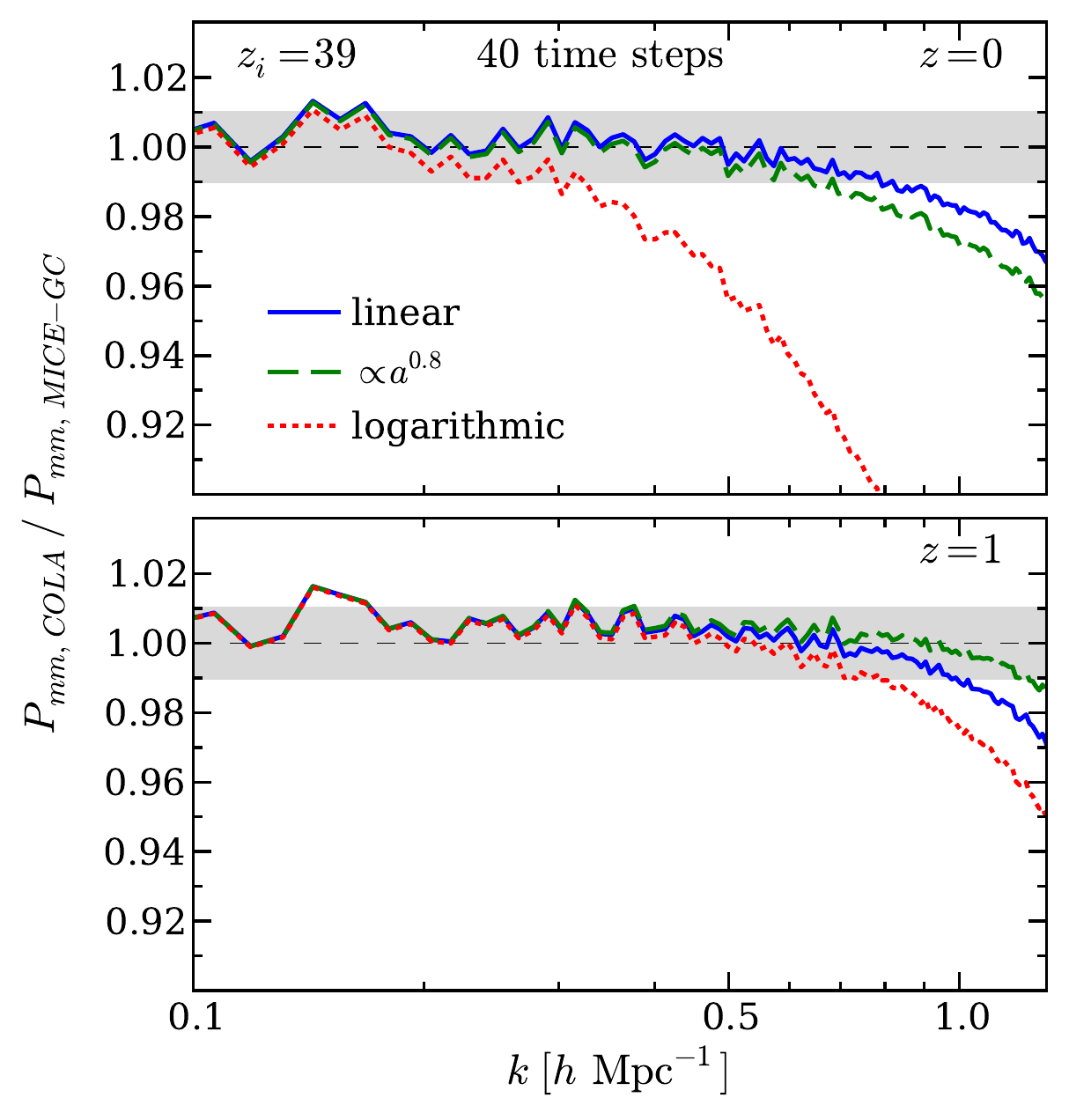}
\caption{Matter power spectrum for the following time sampling functions: linear (solid line), $a^{0.8}$ (dashed) and logarithmic (dotted). All runs contain 40 time steps starting at $z_i=39$. The upper and lower panels show redshifts 0 and 1, respectively. The case that performs better at all redshifts is the linear one.}
\label{fig:pkmm_vs_t-sampling}
\end{figure} 

A logarithmic time sampling is useful in full $N$-body codes for global time steps that affect all particles. But these algorithms in general implement adaptive time stepping schemes for individual particles that sample more accurately the time evolution when non-linearities start to grow. In the implementation of \cola we are using there is no such refinement at late times and we see in Fig.~\ref{fig:time-stepping} that the logarithmic choice oversample the early evolution of the particle distribution at the price of a low step density at later times. On the other hand, the linear case has large relative variations on the scale factor during the first steps, which might presumably lead to larger inaccuracies. For that reason we have considered the power law function to sample intermediate situations if $0 < p < 1$.

In Fig. \ref{fig:pkmm_vs_t-sampling} we compare the matter power spectrum for three time sampling functions: linear, $a^{0.8}$ and logarithmic in solid, dashed and dotted lines respectively. Upper and lower panels correspond to redshifts 0 and 1 respectively. We see that any gain we might have at high redshift by concentrating there more time steps is lost as soon as the step density decreases later. This is evident for the logarithmic case, which has the lowest step density for $z<2$. In particular, at $z=0$ the power spectrum is close to that for the case of 10 time steps linearly distributed (see dotted line in the upper panel of Fig. \ref{fig:pkmm_vs_nsteps}) and indeed they have a similar step density. The distribution of time steps using the function $a^{0.8}$ provides better results at high redshift, but falls below the linear distribution for lower redshifts where it has a lower step density.

Therefore, an optimal distribution should have a step density without strong variations and we conclude from the measurements that the linear case offers the best global performance at all redshifts. Although large relative variations on the scale factor during the first steps could lead to inaccuracies, this is balanced by the fact that at early times the dynamics is close to linear and can be well approximated by the 2LPT evolution. Hence a better time sampling at the beginning is not as critical as for low-$z$. Since all these arguments are built based on relative variations of the step density across time, \emph{the conclusions are independent of the absolute number of time steps and the initial redshift}. Therefore in the rest of the paper we shall adopt a linear time-stepping distribution.

Lastly, we can use the concept of step density to frame the results from Sec.~\ref{sec:nsteps} for the abundance of haloes. We find that whenever the step density is low ($\rho\la20$), the mass function suffers an underestimation for masses above $10^{13.5}\Msun$.

\subsection{Initial redshift}
\label{sec:intial_redshift}

The optimal selection of the initial redshift is coupled with the number of time steps (but it is independent of the time sampling distribution, as we have already shown).

A first guess we can do is to set the initial scale factor equal to the step width, which for 10 time steps gives $z_i=9$ \citep{Tassev13}. Starting later would introduce more transient effects whereas doing it earlier would produce large relative variations in the scale factor in the first time step, which seems not optimal. So there is not much room for optimization using few time steps. Instead we now focus on the situation in which we have 40 time steps.

\begin{figure} 
\includegraphics[width=0.99\columnwidth]{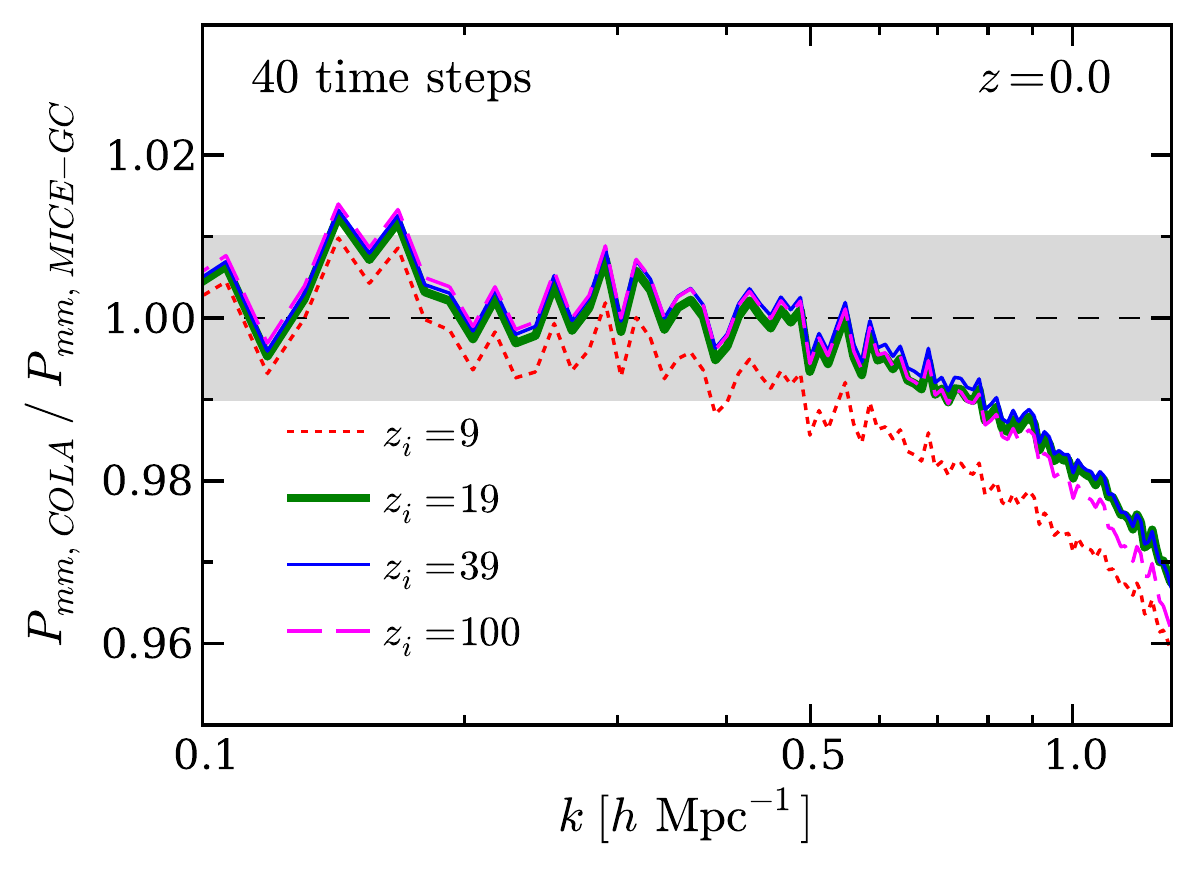}
\caption{Matter power spectrum at $z=0$ as a function of the initial redshift. All runs distribute 40 time steps linearly along the scale factor. A too low starting redshift produces transients at non-linear scales. The cases with $z_i \gtrsim 19$ are almost indistinguishable from each other, except for a slightly less power for $z_i=100$ at $k \sim 1 \hompc$ (see text for details).}
\label{fig:pkmm_vs_zi}
\end{figure} 

Using the same rule as before we can estimate a good guess as $z_i=39$. In Fig. \ref{fig:pkmm_vs_zi} we show the resulting matter power spectrum at $z=0$ when the initial redshift is varied from 9 to 100. A low value of $z_i=9$ yields transient effects at all wave-numbers with an amplitude up to one per cent. The rest of cases are almost indistinguishable, only for $z_i=100$ there is slightly less power at $k\sim1\hompc$. On the other hand, we detect that the mass function is underestimated at low masses if the initial redshift is too high. One possible explanation is that for high starting redshift the density contrast in the initial conditions are too smooth and then, due to the coarse time sampling in {\tt COLA}, the smallest density peaks that seed small mass haloes are blurred. This pushes towards using an initial redshift not as high so fluctuations are larger, even if they are given by 2nd order in LPT.

Nonetheless the dependence we observe on this parameter once $z_i \sim 20-40$ is weak, since in practice only affects the position of the first(s) time step. But given the slightly better performance on the mass function of $z_i=19$ when using 40 time steps, we adopt this value as our fiducial choice in what follows.

\subsection{Force mesh grid size}
\label{sec:force_mesh_grid}

Previous subsections were devoted to parameters that define the temporal discretization of the simulation. We use now the most accurate configuration (that is, 40 time steps linearly distributed from $z_i=19$) to study the effects of the spatial discretization in the force computations. In particular, we compare runs with ${\rm PM}_{grid}$ factors of 2 and 3. We note that this parameter is of particular relevance as it has a large impact in the computational cost of the runs. For example, ${\rm PM}_{grid}=2$ allows a saving of $70\%$ of the computing time and $30\%$ of the memory consumption with respect to ${\rm PM}_{grid}=3$.

\begin{figure} 
\includegraphics[width=0.99\columnwidth]{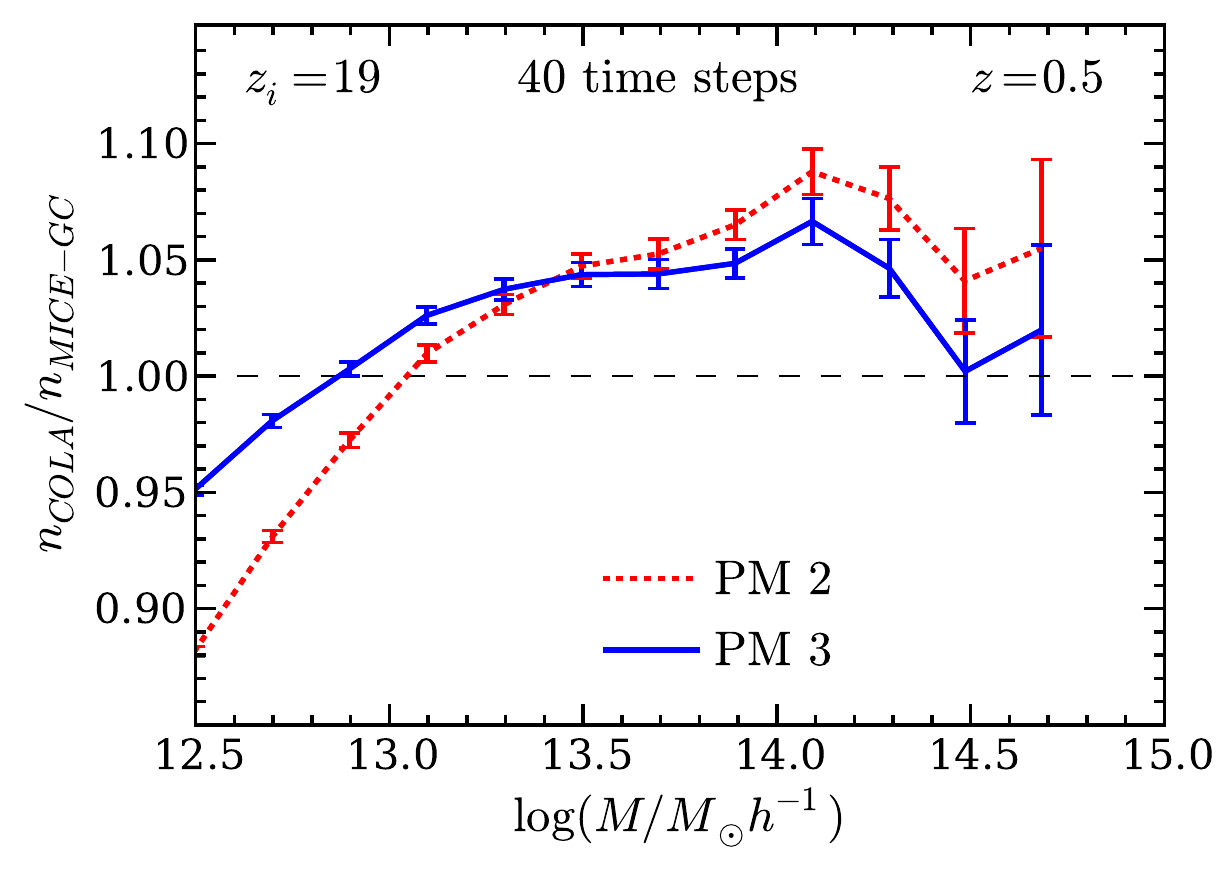}
\caption{Mass function at $z=0.5$ using a ${\rm PM}_{grid}= 2$ and $3$ (dotted and solid lines, respectively). Both runs contain 40 time steps and have $z_i=19$. Decreasing the size of the PM grid produces a larger overestimation of masses for large haloes and increases the incompleteness for small haloes.}
\label{fig:mf_vs_pm}
\end{figure}

We find that changing from ${\rm PM}_{grid}=3$ to $2$ only changes the matter power spectrum by $\sim 1\%$ for $k\la1\hompc$. However there is a more important effect on the halo mass function. For ${\rm PM}_{grid} = 3$, what corresponds to a comoving cell size of $0.25\mpcoh$ given our box-size and particle load, we recover a mass function within $5\%$ of the one measured in MICE-GC down to $10^{12.5} \Msun$. This is shown in Fig.~\ref{fig:mf_vs_pm} with a solid blue line. According to the halo model this mass scale corresponds to a halo size of $\sim 2 \mpcoh$ (e.g. \citealt{2001ApJ...546...20S}). This means that in order to resolve a given halo mass scale at such accuracy we need a minimum of roughly 8 cells to sample its typical halo size. Otherwise the haloes are puffy and might not collapse and be resolved. For ${\rm PM}_{grid}=2$ the force evaluation cell size is $\sim 0.375\hompc$, given the above scaling it means that we should only resolve the mass function at $\sim 5\%$ for haloes more massive than $\sim 10^{13} \Msun$. This is in fact in good agreement with our findings in Fig.~\ref{fig:mf_vs_pm}. 

A discrepancy in the mass function might have two sources: a genuine difference on the abundance or that halo mass estimates are systematically biased. The first case, and assuming that the difference is spatially homogeneous, does not produce differences in clustering for samples selected by mass cuts (see Table \ref{table:mass_bins}), while the second does. Since we do not detect any significant difference in halo clustering at the low mass range for different ${\rm PM}_{grid}$ factors, we infer that there is a completeness problem at those masses due to the size of the force mesh. Not all haloes that should form are detected in the simulation, and in a mass-dependent way.

At high masses, on the contrary, we observe a lower clustering amplitude ($\sim1$ per cent at linear scales) for the run that produces a higher overestimation on the mass function. Both facts can be explained by a halo mass overestimation. One possible interpretation is that the puffier the haloes due to the force resolution, more easily the FoF algorithm bridges neighboring particles or small groups to a halo that really do not belong to it, hence systematically biasing high the mass estimate, as we observe.

\subsection{Optimal setup}
\label{sec:optimal_setup}

So far, we have given an exploration of the main code parameters in \cola and their impact on the dark matter clustering and on halo abundance. To achieve percent accuracy on both quantities, at very least 10 time steps have to be done before the redshift of interest\footnote{Using a particle mass of $~2.9\times10^{10}\Msun$.}, which means that in total we might need 20 or more until $z=0$. The more we do, the higher is the wavenumber where the dark matter power spectrum starts to miss power. This is true to at least $\sim40$ time steps, above that one should probably set ${\rm PM}_{grid}>3$ (for the reference mass resolution we use, $2.9\times10^{10}\Msun$), so that the force resolution does not limit the accurate sampling of power up to $k \sim 1\hompc$. In Appendix \ref{sec:pmonly} we show that still with these configurations, the \cola method yields better results than a PM only evolution.  After we find that the linear time sampling distribution is the optimal one, regardless of the rest of parameters, and that for a large number of time steps best results are already achieved with an initial redshift of 19. A high ${\rm PM}_{grid}$ factor is required for percent accuracy in halo abundance and matter clustering and thus we set ${\rm PM}_{grid} = 3$ despite its relatively higher computational cost. The prize of further increasing it is not well justified in terms of additional accuracy.

A good choice of code parameters depends on which accuracy requirements need to be accomplished by the final mock catalogues. In this work, our target is to achieve per cent level accuracy on the matter power spectrum and halo abundance, in the wide ranges of $10^{12.5}-10^{15.0}\Msun$ in mass, scales up to $k\sim1\hompc$ and redshifts comprised between 0 and 1.5.

\begin{figure} 
\includegraphics[width=0.99\columnwidth]{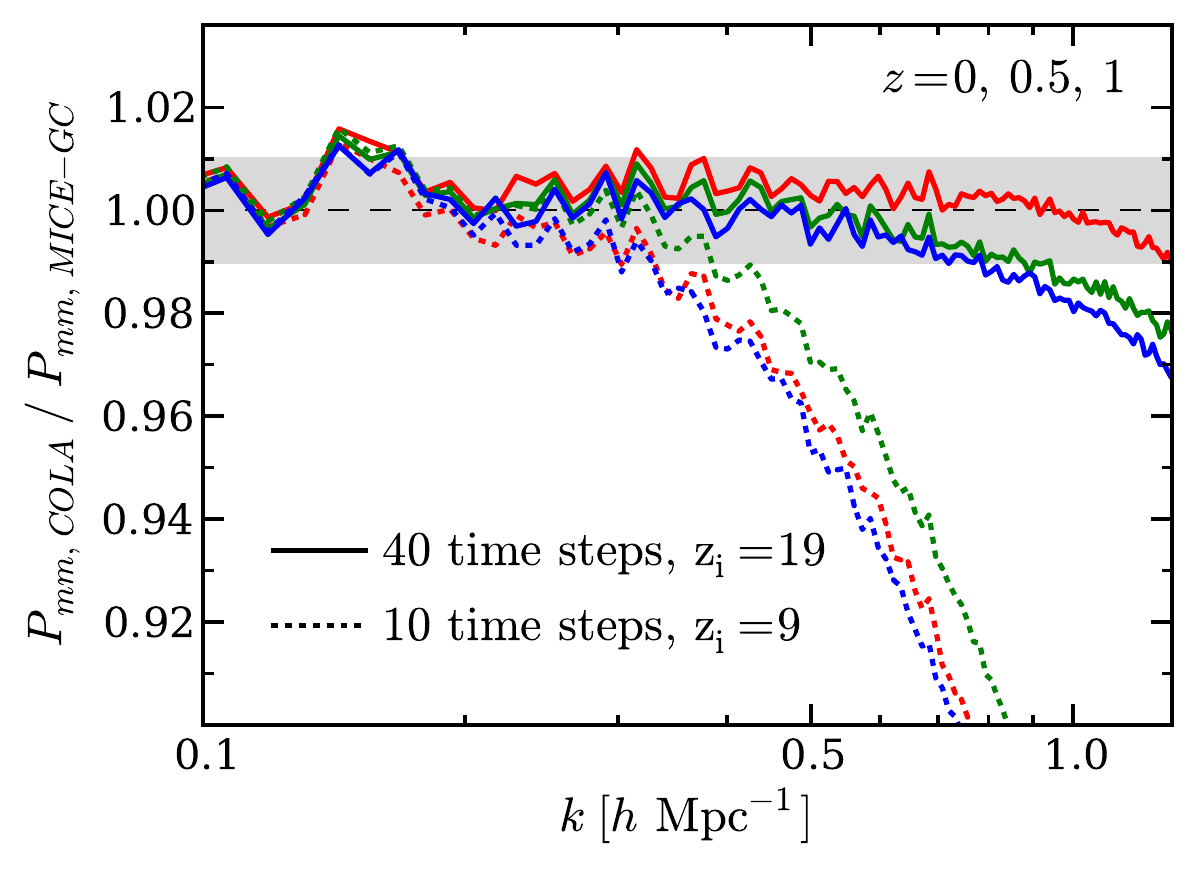}
\caption{Comparison of the matter power spectrum in real space for the run using optimal parameters (40 time steps and $z_i=19$, solid lines) and the default configuration (10 time steps and $z_i=9$, dotted lines). We show redshifts 0, 0.5 and 1. This optimal setup delivers a $\sim1$ per cent accuracy at scales $k\sim1\hompc$.}
\label{fig:pkmm_optimal_vs_10steps}
\end{figure} 

Given these requirements we find that the best setup is set by 40 time steps linearly distributed along the scale factor, starting at $z_i=19$ and with a ${\rm PM}_{grid}$ factor of 3. Fig. \ref{fig:pkmm_optimal_vs_10steps} shows the matter power spectrum using this configuration (solid lines) compared with the case of 10 time steps starting at $z_i=9$ (dotted lines) at redshifts 0, 0.5 and 1. At $z=0$ (1), there is a 1 per cent accuracy up to $k=0.8$ ($1.3$)$\hompc$. Regarding the mass function, the solid line in Fig. \ref{fig:mf_vs_pm} depicts results at $z=0.5$ for the optimal setup. For other redshifts the conclusions are quite robust, i.e., a 5 per cent underestimation at $M=10^{12.5}\Msun$ and a $\sim5$ per cent excess for $M\ga10^{13.5}\Msun$.

\section{Halo clustering}
\label{sec:halo_clustering}

\begin{figure*}
\includegraphics[width=1.99\columnwidth]{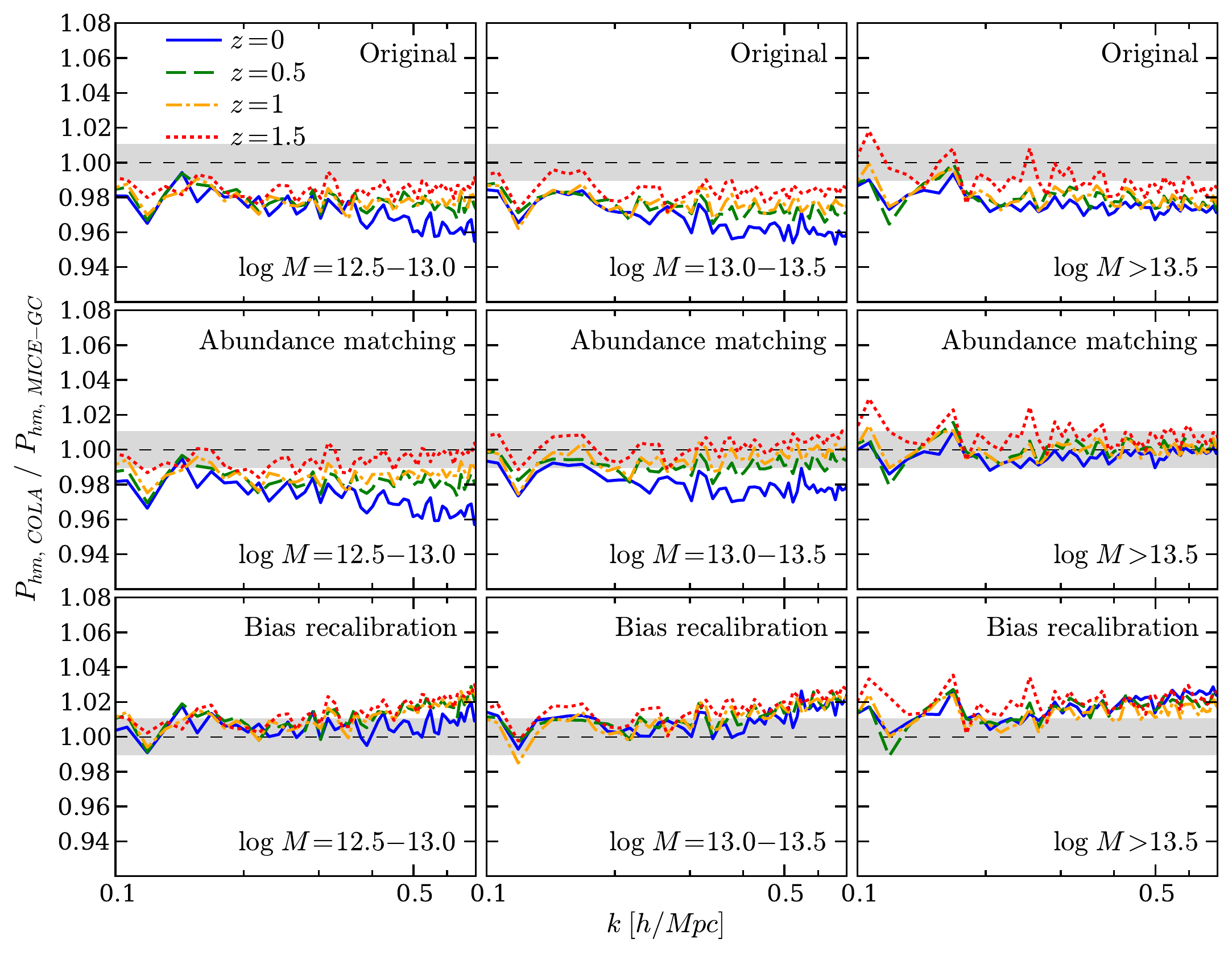}
\caption{Halo-matter cross power spectrum. Rows correspond to different halo mass corrections: original mass, after abundance matching (see Sec. \ref{sec:AM}) and after bias re-calibration (see Sec. \ref{sec:bias_recalibration}) from top to bottom. Columns separate halo samples M1, M2 and M3 from left to right. The original catalogue has a bias underestimation of $\sim2$ per cent in all cases. The abundance matching performs well at high masses while the bias re-calibration is able to achieve a 1 $\%$ agreement on large scales (i.e, $k < 0.3 - 0.4\hompc$)  and 2 $\%$ at intermediate scales ($0.4 < k < 0.7\hompc$).}
\label{fig:pkhm_raw_ham_bias}
\end{figure*}

In Sec.~\ref{sec:optimization} we found an optimal configuration set-up for \cola by benchmarking the matter clustering and the halo abundance as a function of redshift against those measured in MICE-GC. We now study what that configuration implies for the clustering of haloes.
The first row in Fig. \ref{fig:pkhm_raw_ham_bias} shows the halo-matter cross power spectrum in real space and without applying any correction to the catalogues. Different columns separate mass samples M1, M2 and M3 (see Table \ref{table:mass_bins}) from left to right. Solid, dashed, dot-dashed and dotted lines display redshifts 0, 0.5, 1 and 1.5 respectively. The other two rows are explained in the following subsections \ref{sec:AM} and \ref{sec:bias_recalibration}. We notice that there is a general $\sim2$ per cent under-estimation of the clustering amplitude at all mass bins and redshifts. This constitutes a remarkable result: it is possible to predict the halo linear bias with an accuracy of $\sim2$ per cent without doing any correction nor the necessity of calibrating against a reference full $N$-body simulation.

Note, however, that we have also found evidence that halo masses are biased. Hence, we now explore two different corrections on the mass with the aim of reducing further the deviations in the halo bias to the $\lesssim 1\%$ level. One is based on fitting abundances (abundance matching, Sec. \ref{sec:AM}) and the other on fitting clustering (bias re-calibration, Sec. \ref{sec:bias_recalibration}).

\subsection{Abundance Matching}
\label{sec:AM}

The cumulative halo mass function gives a monotonic relationship between the mass and the abundance of haloes. Biased halo mass estimates makes this function in \cola to have deviations with respect to a reference $N$-body simulation. If we have an external fiducial mass function (coming from a full $N$-body simulation for instance), we can re-assign the halo masses in the catalogue so that the reference abundance is fitted. When the incompleteness is negligible, we expect this calibration to greatly reduce disagreements among both catalogues if the ranking of halo masses has the correct ordering. If the incompleteness is present, there are missing entries in the catalogue and trying to match abundances will not produce the desired effect but a mixing of haloes with different clustering properties.

The second row in Fig. \ref{fig:pkhm_raw_ham_bias} shows the halo bias after correcting halo masses by abundance matching, using the measured mass function in MICE-GC as reference. The small disagreements in the top panels are greatly corrected in mass samples M2 at $z>0$ and M3 at all redshifts, but not in M1.
This is  consistent with the impact of incompleteness in the mass sample described above: abundance matching works well as long as the incompleteness is not present, i.e. $M \gtrsim 10^{13}\Msun$ (see the solid line in Fig. \ref{fig:mf_vs_pm}).

We have tested as well the capabilities of the abundance matching for runs using only 10 time steps, in which the ``uncorrected'' mass function is highly under-estimated at $z=1$ and the halo bias deviates by $\sim20$ per cent (see Fig. \ref{fig:10steps}). After abundance matching, the bias is recovered at the 3 per cent level for all mass bins and redshifts, but only for $k<0.5\hompc$, what illustrates that mass calibration performs worse when non-optimal parameters are used in {\tt COLA}.

\subsection{Halo bias re-calibration}
\label{sec:bias_recalibration}

We now explore an alternative mass re-calibration that is targeted to fit the halo bias. Note in the first row of Fig. \ref{fig:pkhm_raw_ham_bias} the \cola run (with optimal set-up) yields always a residual bias mismatch of 2 per cent, regardless of the mass sample and redshift. We can use this fact to build an alternative correction independent of any parent simulation (assuming the $2\%$ factor is roughly independent of cosmology). In the framework of the halo model \citep{Cooray02} halo bias and halo mass are related through a function that only depends on cosmology $b=b(M)$. Thus,
to first order, a fractional reduction in the bias of $\delta \ln b$ can be recovered with a shift in halo mass $\ln M \rightarrow \ln M - \delta \ln M $ given by,
\begin{eqnarray}
\delta \ln M = \left(\frac{\partial b}{\partial \ln M} \right)^{-1} b \, \delta \ln b. 
\label{eq:mass_re-calibration}
\end{eqnarray}

In what follows we set $\delta \ln b=0.02$ (the bias calibration value we found) and evaluate the derivative in Eq.~\ref{eq:mass_re-calibration} at the corresponding mass and redshift using the bias prediction from \citet{Sheth-Tormen99} but we have checked that other fitting functions provide similar results. 

The recovered halo bias values after doing such mass re-calibration are shown in the third row in Fig. \ref{fig:pkhm_raw_ham_bias}. Now the agreement with MICE-GC is within $1\%$ up to scales $k \lesssim 0.5 \hompc$ for all redshifts and masses. However, the correction is not working perfectly for the mass sample M3, where it is sub-percent up to $k \lesssim 0.3 \hompc$ but yields an over-estimate of $\sim 2\%$ beyond. We believe this could be due to the limited accuracy of the bias predictions coming from the theory of the peak background split \citep{Manera10}, used to evaluate Eq.~(\ref{eq:mass_re-calibration}). Provided with a better bias prescription (or maybe the bias-mass relation measured from a reference $N$-body itself) one would expect the bias re-calibration to give very good results by construction. Nonetheless the accuracy remains within $1\%$ for most cases and deviations from that are small.

As was the case for the abundance, this correction solves disagreements due to biased halo masses but not due to incompleteness. The over-abundance at high masses is removed but the underestimation at low masses persists and even increases. Despite that, haloes have the right clustering amplitude.

\section{Redshift space}
\label{sec:z-space}

We now turn into discussing the performance of \cola for reproducing observables in redshift space, as this is what is actually observed in large scale structure galaxy surveys. Redshift space positions $\bmath{s}$ are obtained by,

\begin{eqnarray}
\bmath{s}=\bmath{r}+\frac{\bmath{v}_r}{a H(a)}
\end{eqnarray}
where ${\bmath r}$ is the position in real space, $\bmath{v}_r$ is the peculiar velocity along the line of sight direction, $a$ is the scale factor and $H = a^{-1}da/dt$ the Hubble expansion rate. For concreteness we will focus in halo power spectrum multipoles and assume the plane-parallel approximation, that is, fixing the line of sight to one of the three Cartesian axes. 

In order to reduce the statistical errors in higher order multipoles we have produced 
48 \cola runs using the optimal setup described in Sec. \ref{sec:optimal_setup}. We split the halo catalogue in the 3 mass bins as in Table \ref{table:mass_bins}, with halo masses re-calibrated using the bias method with $\delta \ln b=0.02$ (see Sec. \ref{sec:bias_recalibration}). 

\begin{figure*}
\includegraphics[width=1.99\columnwidth]{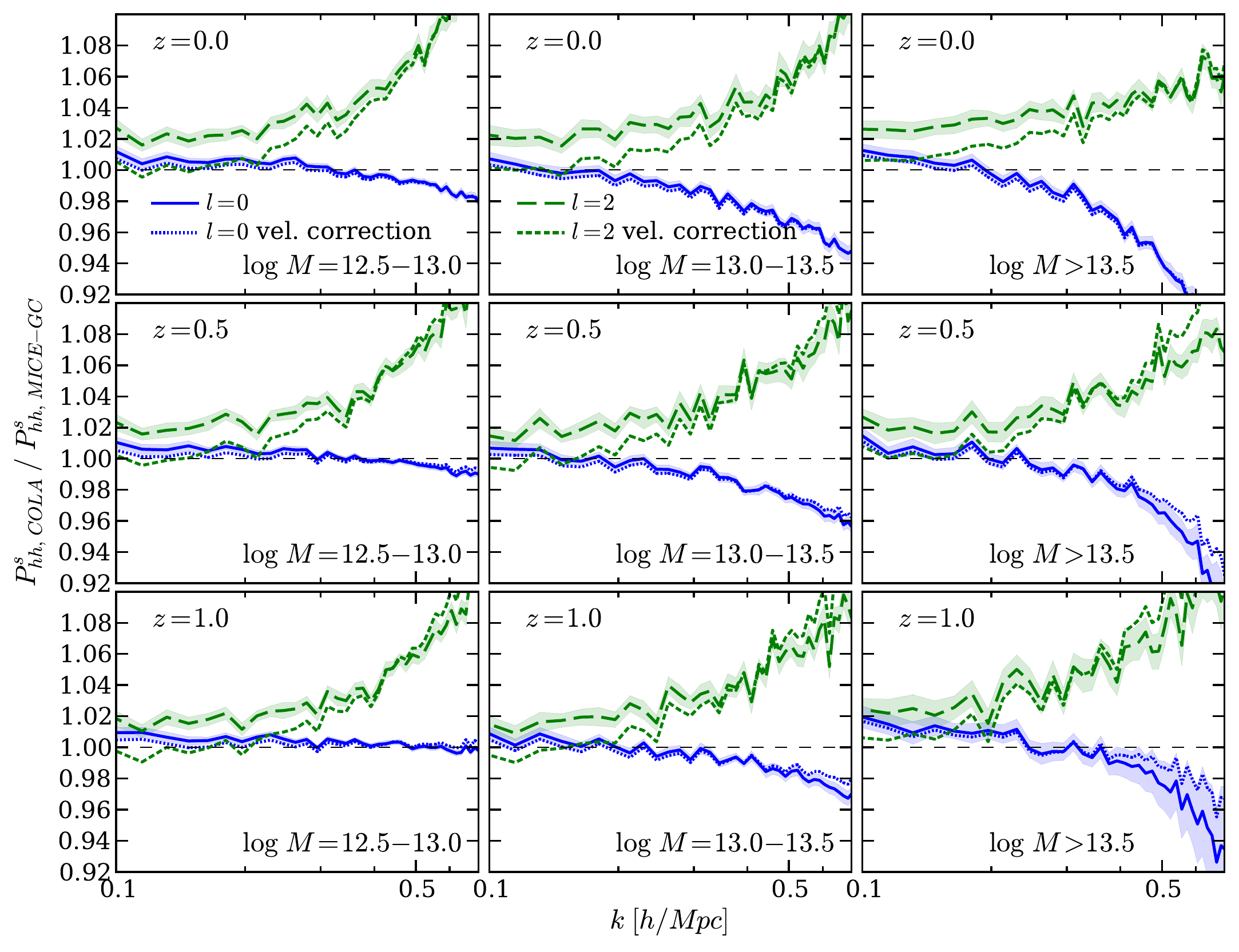}
\caption{Monopole ($l=0$) and quadrupole ($l=2$) of the halo power spectrum in redshift space (solid and dashed lines respectively) in \cola vs. the MICE-GC $N$-body simulation. Different rows correspond to redshifts 0, 0.5 and 1 from top to bottom and columns separate mass samples from left to right (halo masses have been corrected by the bias calibration method). Monopoles have been corrected for shot-noise. Measurements in \cola correspond to the mean over 48 runs using the optimal setup. At large scales ($k<0.3\hompc$) the agreement is within 1 per cent for the monopole and $2\%-3\%$ for the quadrupole. Dotted and short-dashed lines are the monopole and quadrupole after reducing halo velocities by 2 per cent, what brings the latter to better agreement while leaving the former unchanged.}
\label{fig:pkzhh_bias_cal2p}
\end{figure*}

\begin{figure*}
\includegraphics[width=1.99\columnwidth]{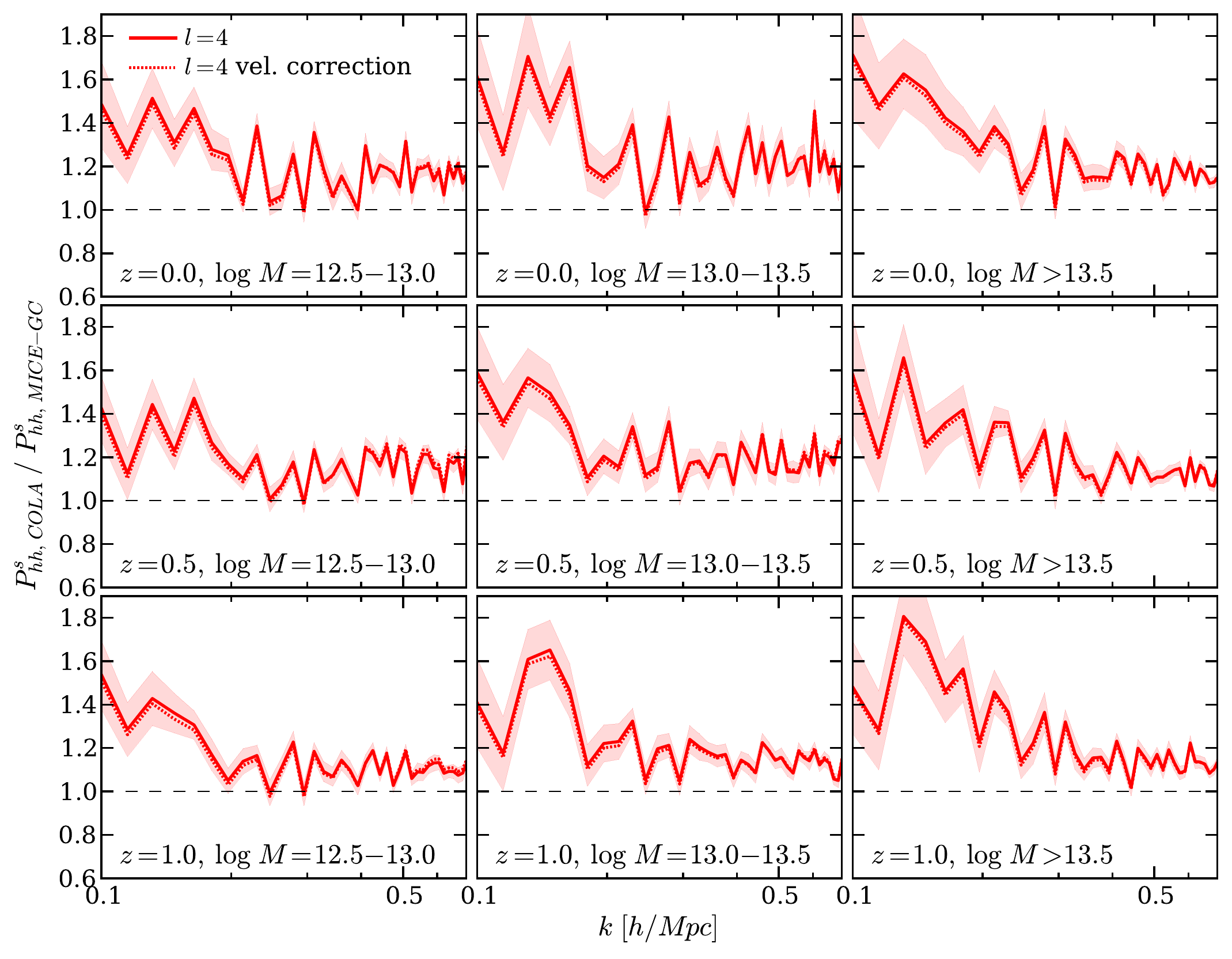}
\caption{Mean halo power spectrum hexadecapole in \cola (after the correction in the halo masses by the bias re-calibration method with $\delta b/b =0.02$) compared to the one in MICE-GC. 
Different rows correspond to redshifts 0, 0.5 and 1 from top to bottom and columns separate mass samples M1, M2 and M3 from left to right. For $k>0.2\hompc$ the agreement with the $N$-body is at the $\lesssim 20\%$ level across redshifts and mass bins. The dotted line includes as well a reduction in halo velocities of 2 per cent (see text for details).}
\label{fig:pkzhh_bias_cal2p_hex}
\end{figure*}

Figure \ref{fig:pkzhh_bias_cal2p} shows the mean of the monopole (l=0) and the quadrupole (l=2) over the suite of \cola runs divided by the corresponding quantities measured in MICE-GC. Rows separate redshifts $0, 0.5$ and $1$ from top to bottom and columns mass bins M1, M2 and M3 from left to right.

For reference we recall the large-scale limit expressions for these quantities \citep{Kaiser87} assuming a simple linear bias model,
\begin{eqnarray}
\label{eq:multipoles}
P^{\mathrm{s}}_{l=0,hh}(k)=&\left(b^2+\frac{2}{3} b f +\frac{1}{5}f^2\right)P^{\mathrm{r}}_{mm}(k), \nonumber \\
P^{\mathrm{s}}_{l=2,hh}(k)=&\left(\frac{4}{3} b f +\frac{4}{7}f^2\right)P^{\mathrm{r}}_{mm}(k) ,
\end{eqnarray}
where $b$ is the bias and $f \equiv\frac{\mathrm{d}\ln D}{\mathrm{d}\ln a}$ the linear growth rate.

At large scales ($k<0.3\hompc$), the agreement with MICE-GC is within 1 per cent for the monopole. Recall that we are using bias-recalibrated masses what ensures that the halo clustering is well reproduced in real space. And this contribution is the leading order for the monopole in redshift space, on large scales (i.e. the $b^2 P^r_{mm}$ term in Eq. \ref{eq:multipoles}). Had we used the actual \cola halo masses instead we would have obtained biases off by 2 per cent and the monopole underestimated by at least $\sim 4\%$. In turn, the quadrupole in Fig.~\ref{fig:pkzhh_bias_cal2p}  is systematically overestimated by $\sim 2$ per cent across all mass bins and redshifts ($k<0.3\hompc$). On large scales, the leading order contribution to the quadrupole is the cross-correlation between halo densities and halo velocities, i.e. the term $b f P^r_{mm}$ in Eq. \ref{eq:multipoles}. This means that any inaccuracies in reproducing the velocity field by \cola will have a direct impact in the quadrupole. For instance we have checked that the differences on large-scales can be corrected by reducing by 2 per cent each halo velocity (what would amount to reduce the overall bulk flow). We over plot (without error bars) the monopole and the quadrupole with dotted and short dashed lines respectively after applying such velocity correction. As expected, the quadrupole is now perfectly in agreement at large scales and the monopole remains almost unaltered.

At smaller scales ($k>0.3\hompc$) we observe larger discrepancies. The monopole is underestimated, specially at high masses, and the quadrupole is overestimated. We believe this is due to the details of the full velocity PDF\footnote{For example, we have measured the halo 1-dimensional velocity distribution and found that the fraction of haloes with center of mass velocities larger than 500 km/s is slightly underestimated by few percent in \cola (the exact number varies for mass samples and redshift), although the halo velocity rms agrees within 1 per cent with MICE-GC.} but we do not attempt to  tune the results further to those of MICE-GC as i) the results on these scales will eventually depend on the galaxy sample under consideration and ii) these are scales that start to be smaller than those used in standard large-scale structure probes such as BAO. For instance, small-scale corrections can be postponed to a later stage when haloes are populated with galaxies using an HOD prescription, in which the velocity dispersion can be fitted to have agreement with observations.
For reference, we have checked that adding a dispersion component to the halo velocities drawn from a Gaussian distribution with a width of $\sim35\,\mathrm{km\,s^{-1}}$ and zero mean reduces the quadrupole for $k>0.3\hompc$ and is then in agreement within 2 per cent for most scales, mass samples and redshifts, whereas the monopole is not substantially affected. 

Figure \ref{fig:pkzhh_bias_cal2p_hex} shows the equivalent of Fig. \ref{fig:pkzhh_bias_cal2p} but for the hexadecapole ($l=4$). We find that  our optimal configuration for \cola  yields an excess of $\sim 50$ per cent at large scales, while for $k>0.2\hompc$ the agreement is significantly improved, down to $\sim 20$ per cent. These differences are not significantly changed when the velocity correction of 2 per cent is applied. If we further add an ad-hoc velocity dispersion term, as discussed above for lower multipoles, we achieve an agreement within 10 per cent at small scales.

\section{Conclusions}
\label{sec:conclusions}

\cola \citep{Tassev13, Tassev15, Howlett15b, Koda15} is a method that represents a step forward in the production of fast and accurate mock catalogues for galaxy surveys.  
It allows speed-ups of almost three orders of magnitude with respect to full accuracy $N$-body simulations (depending on the particular choice of code parameters used) while still reproducing their large-scale structure observables to within few percent.
\cola is based on a combined scheme where particle trajectories are evolved according to LPT with residual displacements to those trajectories computed by a high resolution PM $N$-body solver.

It should be noted that the \cola algorithm can get arbitrarily close to a full $N$-body by demanding more accuracy in the time and space integration, at the expenses of increasing the computational cost. The key is then to find the optimal set-up in the code internal parameter space to balance the highest accuracy in reproducing observables while maintaining low computational requirements. 

In this paper we have determined the optimal configuration of \cola, that we shall name ICE-COLA for reference, by systematically exploring the code parameter space. This optimization uses as target observables the two-point clustering of matter and the abundance of haloes across a wide range in halo mass and cosmic time.
For this purpose, we have developed a suite of $2048^3$ particles \cola simulations that sample the internal code parameter space and compared such observables to a reference state-of-the-art $N$-body simulation, the MICE-GC. For the optimal set-up we then studied the halo clustering in real and redshift space. The most relevant parameters were the number of time steps of integration before the desired redshift output and the spatial resolution of the force computation.

\begin{description}
\item[\textit{Number of time steps}.] It has a direct impact in the minimum well resolved scale. Integrating 40
time steps yields a matter power spectrum with one percent accuracy up to $0.8\hompc-1.3\hompc$ in the range $0 < z < 1$. This is a large improvement over the standard 10 time steps that yields such accuracy only for $k < 0.3 \hompc$. In turn,
halo masses (and mass function at a given mass) are largely underestimated if less than $\sim 10$ time steps have passed by, especially for massive haloes ($M>10^{13.5}\Msun$). Increasing to 40 time steps we find the mass function to agree with MICE-GC to within $5\%$ for $M>10^{12.5}\Msun$ and $0 < z < 1$ (assuming ${\rm PM}_{grid}=3$, see below). Therefore, for an accurate prediction of matter clustering and halo abundance across cosmic time, one needs to increase the default 10 time-steps by a factor of a few. Above 40, however, further gains might be limited by the force resolution. We show in Appendix \ref{sec:pmonly} that, at least up to 40 time steps, the \cola method is still preferred over a PM only simulation.
\newline

\item[\textit{Force mesh grid size}.]\footnote{Recall that we denote the grid size by ${\rm PM}_{grid}$ and its value is given in units of the number of particles to the one third.} 
We find that in \cola there is an incompleteness in abundance at low halo masses (at $\lesssim 300$ particles per halo, 
that corresponds to $M \lesssim 10^{13}\Msun$ for our reference mass resolution). 
This systematic effect can be mitigated by increasing the force resolution. For our optimal set-up with ${\rm PM}_{grid}=3$ we find the incompleteness at the $5\%$ level, see Fig.~\ref{fig:mf_vs_pm}.  Using a coarser grid of ${\rm PM}_{grid}=2$ reduces significantly the computational cost but
increases the incompleteness by a factor of two in the same mass range.
In turn, the matter power spectrum is almost unaffected for $k\la1\hompc$. 
For an optimal balance of good accuracy and computational cost, we encourage using ${\rm PM}_{grid}=3$.
\end{description}

In addition to the above, we have also explored the \textit{time sampling distribution} and found that a linear distribution along the scale factor gives the best global performance between redshifts 0 and 1 (see Fig. \ref{fig:pkmm_vs_t-sampling}) regardless of the number of steps and the initial redshift. Lastly, transients effects from the
\textit{initial redshift} were found negligible once $z_i\gtrsim 19$ within \cola accuracy. Starting earlier had no clear benefits.

Such an optimal set-up yielded halo bias within $2\%$ to that in MICE-GC up to $k \sim 0.7\hompc$. In redshift space, monopole and quadrupole agree to $< 4\%$ to those in the $N$-body ($k \lesssim 0.4\hompc$). These conclusions hold for all redshifts and mass bins investigated, and do not rely on any re-calibration against an external $N$-body, as it is usually done for \cola \citep{Kazin14,Ross15,Howlett15,Leclercq15}. Note that other methods for producing mock catalogues also depend to a certain degree on full $N$-body simulations for calibrations \citep{Monaco02,Monaco13,Scoccimarro02,Manera15,Coles91,White13,Kitaura15,Chuang15b}. 

We have further improved the accuracy of \cola by investigating two particular recalibration schemes, using the MICE-GC as a reference:

\begin{description}
\item[\textit{Abundance matching}.] As shown in Fig. \ref{fig:pkhm_raw_ham_bias}, it provides percent level accuracy in halo clustering at high masses for the run with optimal parameters. At low masses where incompleteness is present, it has no effect and a 2 per cent disagreement persists. 
The mass function is correct by construction.
\newline

\item[\textit{Halo bias re-calibration}.] This method yields percent level agreement in halo bias at all mass bins and redshifts for the run using optimal parameters (see bottom row in Fig. \ref{fig:pkhm_raw_ham_bias}). Using a better halo bias prediction could improve the correction for the largest mass bin even further. The abundance is corrected at high masses but at low ones ($M \lesssim 10^{12.5}\Msun$) it remains under-predicted.
\end{description}

Using the halo-bias recalibration we find the monopole and the quadrupole have an excess of 1 and 2 per cent, respectively,
on large-scales. The hexadecapole also features an excess of power $\lesssim 20\%$ for $k>0.2\hompc$ where signal to noise is higher, and deviates more for larger scales. Correcting down bulk velocities by 2 per cent provides a 1 per cent agreement at large scales ($k \sim 0.1\hompc$) for both the monopole and the quadrupole. 

At the final stage of writing this paper, \citet{Howlett15b} presented a similar analysis using the code \textsc{l-picola}, which is an alternative parallel implementation of {\tt COLA}. Our work has been independently developed from theirs and the scope of our analysis is rather different and complementary. In particular, they choose coarser and faster parameter configurations. For example they use the same number of mesh cells as particles or less (i.e., ${\rm PM}_{grid} \leq 1$). This reduces substantially the memory and cpu-time requirements, but we have shown that this introduces important systematic effects in the halo mass and abundance. 
In addition, we also considered halo clustering both in real and redshift space. 
\citet{Koda15} also presented their methodology for generating galaxy mock catalogues using {\tt COLA}. They show results in both real and redshift space with ${\rm PM}_{grid}=3$, but they stick to the fiducial case of 10 time steps and do not explicitly explore the code parameter space.

In this paper we have shown how an optimal choice of \cola code parameters, plus a minimal halo mass recalibration, can yield clustering results to per cent agreement with respect to full $N$-body for all mass bins, scales and redshifts
of interest for the new generation of galaxy surveys.

\section*{Acknowledgments}

We thank Darren Reed for running the simulations used in Appendix \ref{sec:transients_correction} and Jun Koda for making his parallel implementation of \cola publicly available.

Simulations in this paper were run at the MareNostrum supercomputer - Barcelona Supercomputing Center  (BSC-CNS, \texttt{www.bsc.es}), through grants AECT-2014-2-0013, 2014-3-0012, 2015-1-0013 and 2015-2-0011. Funding for this project was partially provided by the Spanish Ministerio de Ciencia e Innovacion (MICINN), research projects AYA-2012-39559, AYA-2012-39620, and project SGR-1398 from Generalitat de Catalunya. AI is supported by the JAE program grant from the Spanish National Science Council (CSIC). MC has been partially funded by AYA2013-44327 and acknowledges support from the
Ramon y Cajal MICINN program. PF is funded by MINECO, project ESP2013-48274-C3-1-P and ESP2014-58384-C3-1-P.



\bibliographystyle{mnras}
\bibliography{fast_mocks_paper}


\appendix

\section{Transient effects correction in MICE-GC}
\label{sec:transients_correction}

The MICE-GC simulation used first order LPT initial conditions (Zeldovich approximation) at $z_i=100$. 
This approximation is known to yield transient effects in the distribution of matter and haloes mainly depending on the actual initial redshift used  \citep{1998MNRAS.299.1097S,Crocce06}. In this appendix we quantify these effect on the matter power spectrum and the halo mass function by using first or second order LPT for the concrete configuration of MICE-GC. These differences are then corrected in MICE-GC whenever we compare any measurement against those in {\tt COLA}, because otherwise it would be a source of a systematic error. We estimate them using additional \textsc{gadget-2} simulations using the same cosmology and starting redshift as in MICE-GC, see Sec. \ref{sec:simulations}, and evolving $1024^3$ particles using either first or second order LPT. The box size is $768\mpcoh$ in order to keep the same mass resolution as in the reference MICE-GC $N$-body simulation. For the mass function we use as well a larger box size of $3072\mpcoh$ in order to have good statistics at the high mass end.

\begin{figure}
\includegraphics[width=0.98\columnwidth]{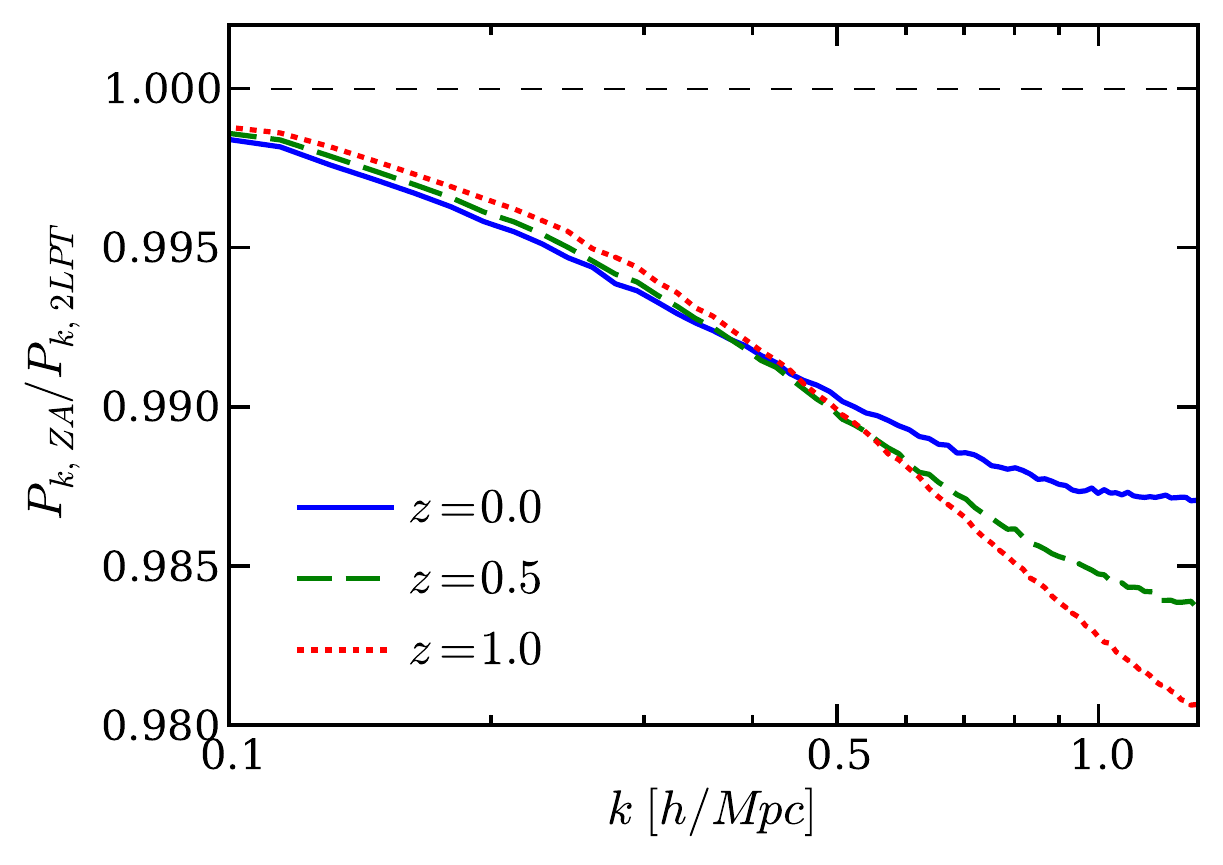}
\includegraphics[width=1.02\columnwidth]{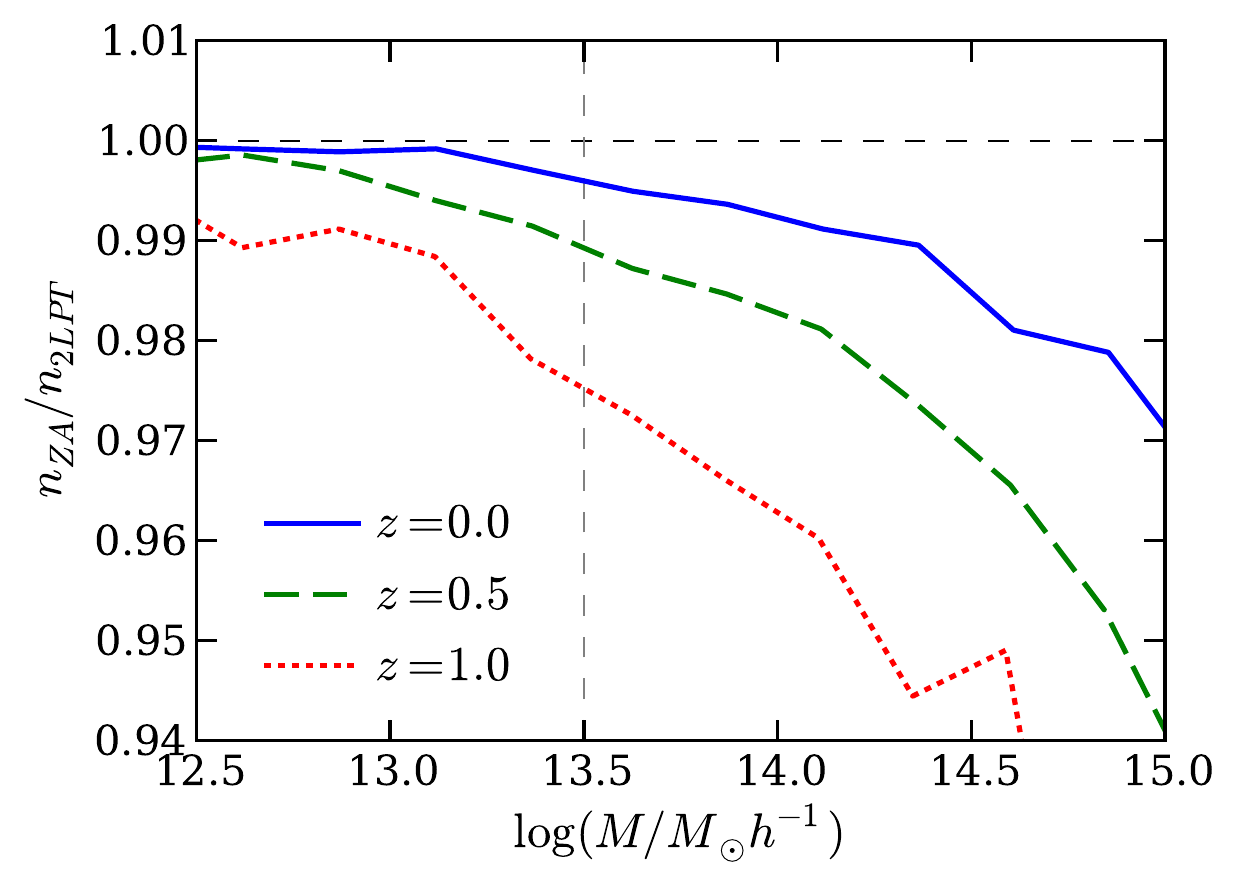}
\caption{Transient effects on the matter power spectrum (upper panel) and the mass function (lower panel). Solid, dashed and dotted lines correspond to redshifts 0, 0.5 and 1 respectively. We show the ratio of the observables measured on a pair of identical full $N$-body runs differing only in the set up of the initial conditions: first order versus second order LPT. The simulations used the same particle mass as MICE-GC in a box of $768\mpcoh$, except for the mass function plot that for $M>10^{13.5}\Msun$ (marked by a vertical dashed line) used a larger box of $3072\mpcoh$.}
\label{fig:pk_transients}
\end{figure}

The upper panel of Fig. \ref{fig:pk_transients} shows the transient effects in the matter power spectrum in real space for redshifts 0, 0.5 and 1 in solid, dashed and dotted lines respectively. The correction is always below 2 per cent in the scales studied and remarkably similar to the results that found by \citep{Schneider15} using another simulation code (see their fig. 2). The lower panel displays the mass function and uses the same line styles. The vertical line at $M=10^{13.5}\Msun$ marks the matching mass-scale for the two runs used (the smaller one for smaller masses and the other way around). Remind that halo masses are defined using the Warren correction and this enables a good overlapping of measurements at that matching mass-scale (in agreement with other tests for such correction, e.g. \citealt{Crocce10}). In the mass function, differences are more important at high masses and redshifts, going up to 5 per cent at $z=1$ for the mass range of interest and they are within 3\% at $z=0$. We measured as well the correction for the halo-matter cross power spectrum, but we found it to be always within the 1 per cent so that the effect is negligible. Thus, whenever we show a ratio of either mass functions or matter power spectra with respect to MICE-GC, we have multiplied it by the corresponding ratio shown in Fig. \ref{fig:pk_transients}. In the case of halo clustering observables we find that transient effects are below the 1 per cent level, so we consider that we can neglect the correction for those measurements.

\section{Performance of \cola with respect to Particle-Mesh only runs}
\label{sec:pmonly}

\citet{Tassev13} showed that \cola simulations with as few as ten time steps recover the matter density field much better than just doing either a particle-mesh (PM) only simulation with the same number of time steps or a 2LPT evolution. In this paper we advocate the use of more time steps in order to produce mock catalogues that are accurate in a large span of redshifts. After increasing the number of time-steps one might think that the 2LPT part of the \cola method has a very little contribution to the dynamics and much of the information comes from the PM integration. In this appendix we show the relative impact of the 2LPT information when many time steps are used.

\begin{figure}
\includegraphics[width=0.98\columnwidth]{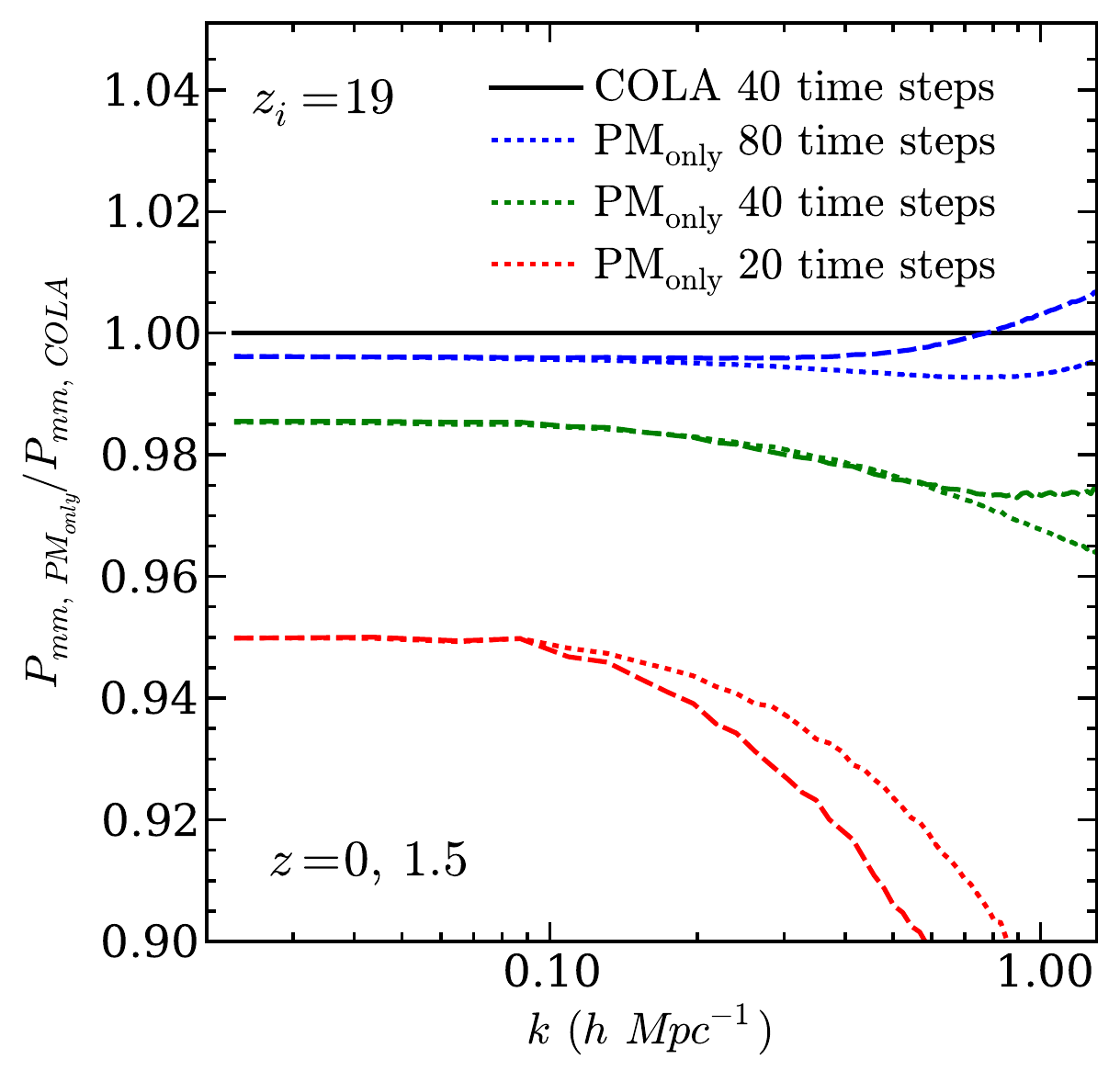}
\includegraphics[width=1.02\columnwidth]{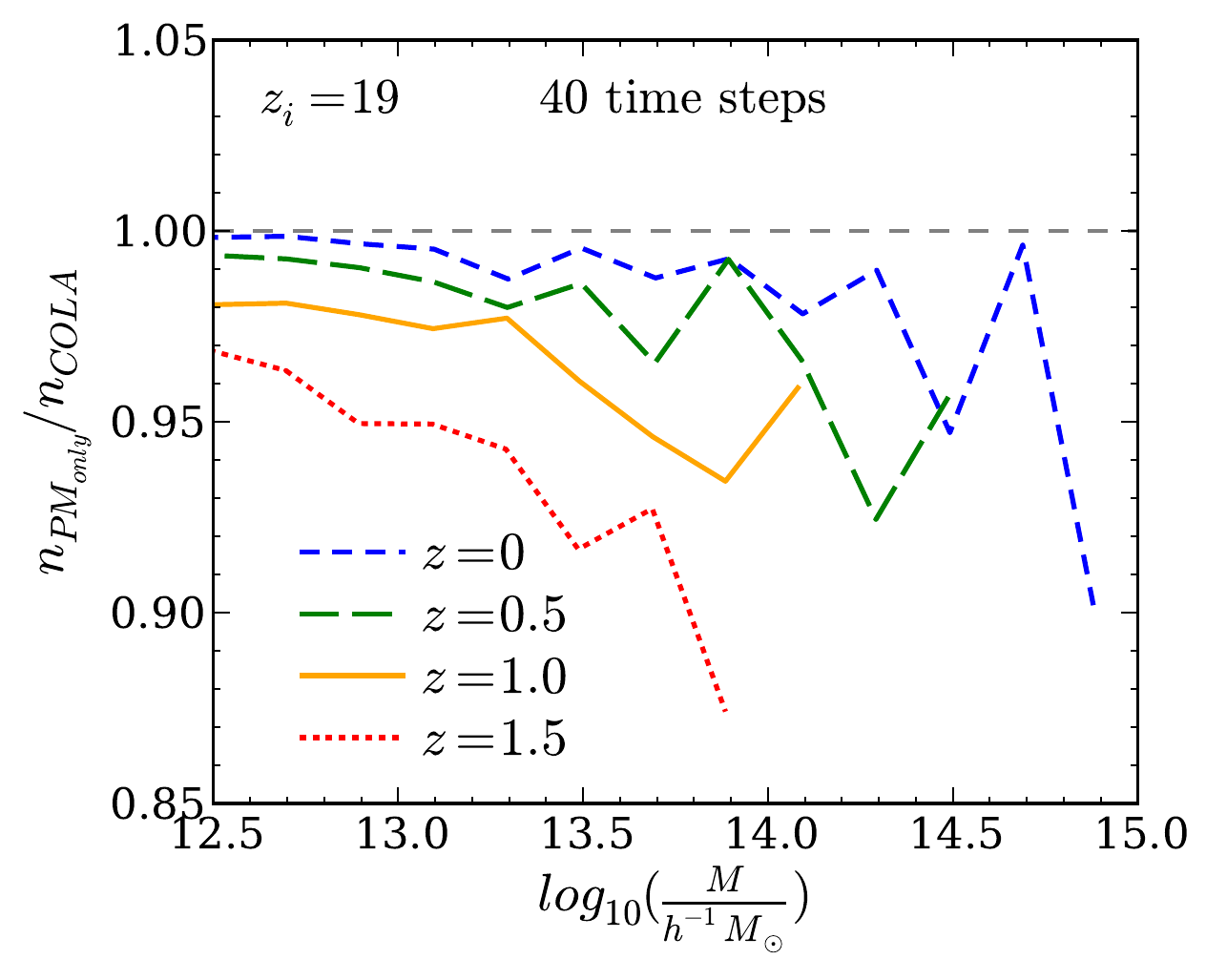}
\caption{{\it Top panel}: matter power spectra for three PM only simulations with 20, 40 and 80 time steps from bottom to top. The reference simulation is a \cola run with 40 time steps. Dotted and dashed lines correspond to redshifts 1.5 and 0 respectively. {\it Bottom panel}: mass function of a PM only run with respect to a \cola one, both with 40 time steps. Plain PM simulations introduce additional systematic effects, even with as much as 40 time steps.}
\label{fig:pmonly}
\end{figure}

For this exercise we use the FastPM\footnote{\texttt{https://github.com/rainwoodman/fastPM}.} parallel implementation of \cola. We run several PM only and \cola runs with $768^3$particles in a box of $576\mpcoh$ by side, and we vary the number of time steps for the PM only runs (the initial redshift is fixed at $z_i=19$). The green line in the upper panel in Fig. \ref{fig:pmonly} shows that the PM only method recovers less power in the matter power spectrum than \cola for the same number of time steps. The deficit is larger at small scales and at high redshift (dashed and dotted lines correspond to redshifts 0.0 and 1.5). The plausible explanation is that the PM only has more difficulties to accurately integrate the equations of motions at high redshifts, when few time steps sample each e-fold of the growth of structures\footnote{Note that in those runs, time steps are linearly distributed with the scale factor.}, and differences persist until $z=0$. The PM method slowly converges to \cola at large scales by increasing the number of time steps. In turn, we recall that \cola reproduces the linear growth rate accurately regardless of the number of time steps (see Fig. \ref{fig:pkmm_vs_nsteps}). The lower panel in Fig. \ref{fig:pmonly} displays the ratio of mass functions between the PM only and \cola runs with 40 time steps, for various redshifts.

There is a clear underestimation that reaches $5-10\%$ for both high redshifts and high masses. We have also studied the halo linear bias and found a corresponding excess exhibiting the same trends. Both differences can be explained by a systematic underestimation of the halo masses for plain PM simulations that is mass and redshift dependent. Both panels of Fig. \ref{fig:pmonly} show that, for a similar number of time-steps, differences between PM only and \cola decrease towards lower redshifts. As shown by the matter power spectrum, discrepancies originate at high redshift, and late non-linear evolution masks them (in a similar fashion than transient effects showed in Appendix \ref{sec:transients_correction}). We do not show in Fig. \ref{fig:pmonly} full $N$-body values since we want to focus only on the relative effect of both methods. Also, it is clear that plain PM simulations converge to \cola in the limit of a large number of time steps.

One might argue that this effect on the mass function could apparently solve the overestimation studied in Sec. \ref{sec:force_mesh_grid}. But this seems just a cancellation of errors that might introduce even more undesired systematic effects. We conclude that it is worth using the \cola method even with as many as 40 time steps. Less time-steps (e.g., 20 or fewer) still produce accurate results for {\tt COLA}, while plain PM simulations show non-negligible biases.

We note that upon resubmission of this paper, \cite{Feng16} presented FastPM and showed results that are consistent with our work.

\end{document}